# Chapter 15: The dynamics of eruptive prominences*

**Nat Gopalswamy**
NASA Goddard Space Flight Center, Code 671, Greenbelt, MD 20771, USA

Abstract
This chapter discusses the dynamical properties of eruptive prominences in relation to coronal mass ejections (CMEs). The fact that eruptive prominences are a part of CMEs is emphasized in terms of their physical association and kinematics. The continued propagation of prominence material into the heliosphere is illustrated using in-situ observations. The solar-cycle variation of eruptive prominence locations is discussed with a particular emphasis on the rush-to-the-pole (RTTP) phenomenon. One of the consequences of the RTTP phenomenon is polar CMEs, which are shown to be similar to the low-latitude CMEs. This similarity is important because it provides important clues to the mechanism by which CMEs erupt. The non-radial motion of CMEs is discussed, including the deflection by coronal holes that have important space weather consequences. Finally, the implications of the presented observations for the modeling CME modeling are outlined.

## 1. Introduction

Prominence eruptions (PEs) describe the process by which a previously quasi-stationary prominence erupts and partly or wholly leaves the Sun. When the eruption happens on the disk, it is referred to as a filament eruption. The prominence visible in an instrument's field of view (FOV) in its moving phase is known as an eruptive prominence (EP). We also use the term prominence eruption (PE) as a synonym for EP akin to the usage of coronal mass ejection (CME) to denote the ejected material. The disappearance of a solar filament (DSF) from the observing pass band (usually in H-alpha) is also referred to as disparition brusque (DB). Filaments may also disappear due to local heating, but this chapter does not concern with such thermal DBs. Prominence eruptions have been known for a long time (see e.g. Kleczek 1964; Martin 1973; Engvold 1980). Kleczek (1964) published a catalog of eruptive prominences occurring between 1938 and 1961. Engvold (1980) provided a detailed discussion on the kinematics, occurrence rates, and source regions of eruptive prominences. Prominence/filament observations exist for more than a century, so there is extensive literature covering PEs. On the other hand, complete CME observations are available only for the past few decades. Therefore we focus only on those aspects PEs that involve CMEs because we now know that PEs are an integral part of CMEs (see e.g., Hildner et al. 1975; Schmahl and Hildner, 1977; Gosling et al. 1976; Hundhausen 1993; Gilbert et al. 2000; Hori and Culhane 2002; Gopalswamy et al. 2003a; Schrijver et al. 2008; Liu et al. 2012; Parenti 2014). Other chapters in this volume by Gibson (2014), Lugaz (2014), Webb (2014), and Fan (2014) provide complementary information on various aspects of eruptive prominences.

## 2. Prominence Eruptions, CMEs, and Flares

Historically, flares and PEs have been known since the nineteenth century. When CMEs were discovered, it was natural to compare PEs with flares and CMEs. In this section we would like to point out that the three processes can hardly be separated. In order to show the interconnection, we start with an example. Figure 1 shows a long east-west filament erupting from the northwest quadrant resulting in a two-ribbon flare and an extended post-eruption arcade (PEA). The eruption was observed by the Extreme-ultraviolet Imaging Telescope (EIT) on board the Solar and Heliospheric Observatory (SOHO) mission. When the filament reached the field of view of the Large Angle and Spectrometric Coronagraph (LASCO) on board SOHO, it was found to be in the interior of a large CME. The angular width of the CME in the sky plane was ~90° and the speed was relatively high (~890 km/s). These values are above average for CMEs observed by most coronagraphs (see e.g. Gopalswamy, 2004 and references therein). The GOES soft X-ray flare size was only B5.0, which means that the flare was rather weak but can be seen clearly above the background as a gradual event for ~6h (the flare size is denoted by the letters A, B, C, M, and X in the increasing order of peak soft X-ray flux by an order of magnitude: A1.0 = 1.0 x $10^{-8}$ $Wm^{-2}$). In EUV images, the PEA remained above background for many hours. There was no active region in the vicinity of the flare, so this is purely a quiescent filament. Yet, it was associated with both a solar flare and a significant CME.

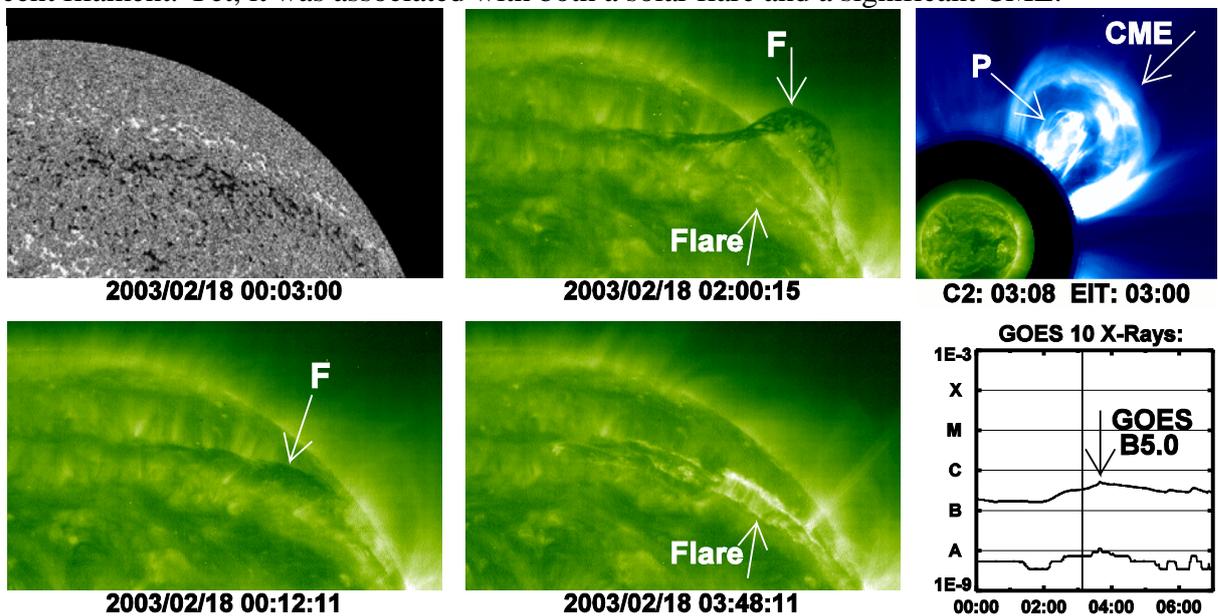

Figure 1. (left) A SOHO/MDI magnetogram (2003 February 18 00:03:00 UT) showing the large-scale bipolar magnetic region and a EUV filament (F) overlying the polarity inversion line (00:12:11 UT). (middle) The filament erupts (02:00:16) accompanied by a flare arcade observed by SOHO/EIT (03:48:11 UT). (right) The associated CME (03:00 UT) with prominence core (P) and the GOES soft X-ray light curve showing a weak flare (B5.0).

### 2.1 Statistical Associations

A high degree of association between PEs and CMEs was recognized soon after the discovery of CMEs (Munro et al. 1979; Webb and Hundhausen, 1987; St. Cyr and Webb, 1991). Munro et al. (1979) found that 70% of CMEs are associated with PEs. The result was similar in studies starting with PEs and connecting them to CMEs (Gilbert et al. 2000; Hori and Culhane, 2002; Gopalswamy et al. 2003b): 72% of PEs were associated with CMEs, when all events automatically detected (Shimojo et al. 2006) from the Nobeyama Radioheliograph (NoRH) images were used. When PEs with radial trajectories were used, the association between PEs and CMEs increased to 83%. A closer examination of the PEs without CMEs revealed that the PEs generally had transverse (parallel to the solar surface) trajectories, or they were stalled while moving in the radial direction. These PEs were the slowest and attained the lowest height (~1.2 Rs from the Sun center on the average). There were also intermediate cases in which transverse PEs attaining slightly larger heights and stalled radial eruptions resulting in detectable changes in the overlying streamers (Gopalswamy et al. 2004a). Some of these streamer-change events may indicate weakening of the pre-eruption configuration because they were followed by PEs and CMEs from the same region. The failed eruptions (Moore et al. 2001; Ji et al. 2003; Guo et al. 2010) are likely to be the "stalled radial PEs".

Small-scale energy release often takes place as a precursor to filament eruptions in the form of compact heating observed in EUV and X-rays (Gopalswamy 1999; Chifor et al. 2007; Sterling et al. 2011a) or nonthermal particles inferred from compact radio bursts (Marque et al. 2001). These signatures indicate reconnection-favoring flux emergence and/or cancelation in the vicinity of filaments that lead to tether cutting (Feynman and Martin 1995; Wang and Sheeley 1999; Chen and Shibata 2000; Gopalswamy et al. 2006). A good example was presented in Gopalswamy et al. (2006). It must be noted that filament eruptions do occur without flux emergence, so there must be other ways in which the filament with its overlying structure gets destabilized and erupts (Schmieder et al. 2013; Aulanier 2014).

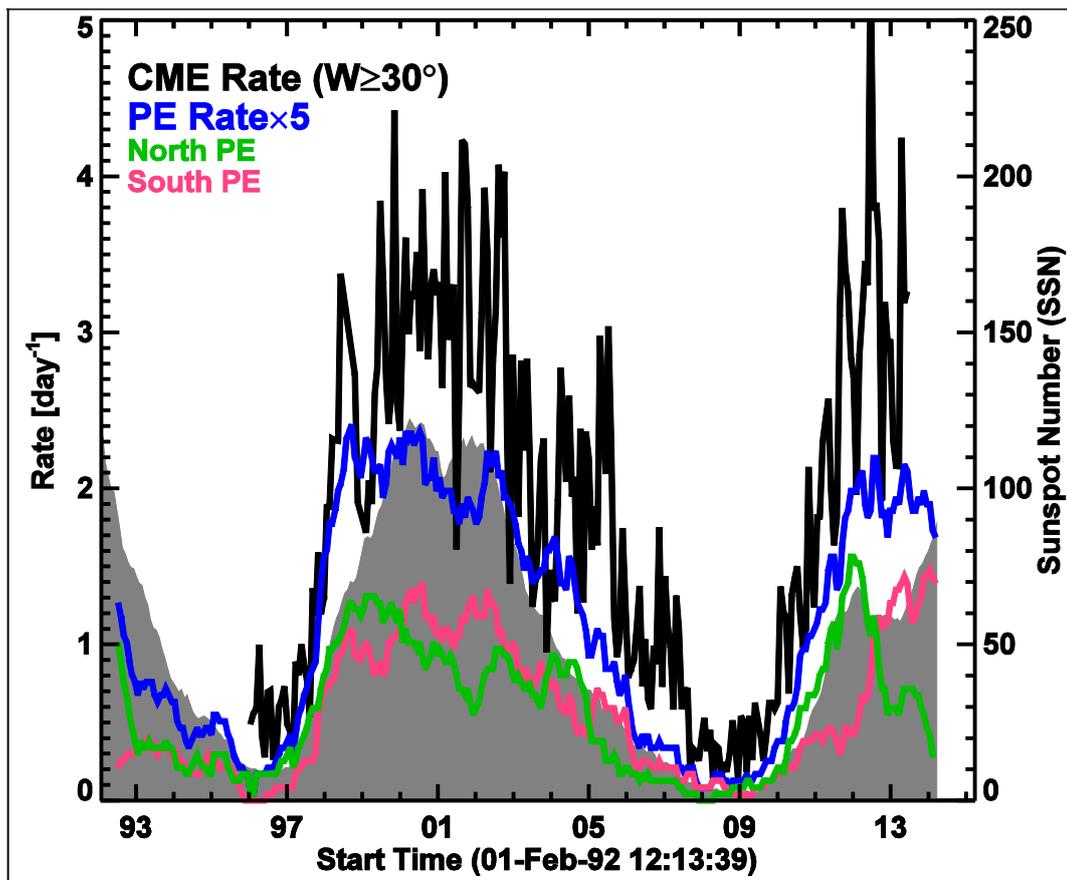

Figure 2. Solar-cycle variation of CME rate (black), PE rate (blue), and sunspot number (SSN - gray). PE rates from the northern and southern hemispheres are distinguished by the green and pink curves, respectively. The start time coincides with the start of operations of NoRH, which is used for the automatic detection of PEs. The PE daily rate is computed as follows. The observed number of PEs in each Carrington Rotation period is multiplied by a factor of three to account for NoRH duty cycle (~8 h per

day). The resulting number is divided by 27.3 days to get the daily rate. The daily rate is then multiplied by a factor of 5 to bring into the scale of the Figure. The CME rate is averaged over Carrington rotation periods.

**2.2 Solar Cycle Variations**

An overall correlation between the variation of PEs and CMEs over solar cycles has been reported earlier (Webb and Howard 1994). Figure 2 presents a long-term comparison of the three manifestations of solar activity: PE rate, CME rate, and sunspot number (SSN). There is a clear drop in SSN between cycles 23 and 24, indicating that the cycle is weak. On the other hand, the PE rates are roughly the same between the two cycles, very similar to what is observed in the CME rates (see also Shimojo 2013;Gopalswamy et al. 2014). The PE rates show a clear north-south asymmetry, with the activity peaking first in the northern hemisphere and then in the south during cycle 23 with a similar trend in cycle 24. A similar asymmetry has also been reported in SSN (Svalgaard and Kamide 2013). The PE rate has a closer similarity to the CME rate than to SSN. This is consistent with the fact that PEs are the most common CME-associated phenomenon at the Sun (Munro et al. 1979). Hundhausen (1993) emphasized the tighter association between "larger-scale" activity such as filaments and helmet streamers on the one hand and CMEs on the other mainly based on similar latitudinal distribution and the long-term variation of that distribution. Hundhausen discounted the importance of "smaller-scale" phenomena such as sunspots, flares, and active regions for CMEs. However, it is necessary to point out that both these large- and small-scale features represent closed magnetic field regions, which can produce CMEs if free energy (the energy available for powering the eruption) can be stored in them. In fact, the most energetic CMEs originate mainly from active regions because large amounts of energy can be stored in active regions. The amount of free energy is roughly given by the potential field energy (volume x $B^2/8\pi$) (Mackay et al. 1997), which can be very large in active regions because of the high magnetic field strength (B). The special populations of CMEs that have significant space weather implications generally originate from the active region belt (see e.g. Gopalswamy et al. 2010a). Filaments are part of active regions too. Active region filaments are thin and short, but can attain much higher speeds similar to the CMEs. The famous backside solar energetic particle (SEP) event with a ground level enhancement (GLE) in cycle 23 on 2001 April 18 was produced by a fast CME (~2500 km/s) and the prominence core had a speed of ~1650 km/s (Gopalswamy, 2006a; Gopalswamy et al. 2012a). Filament eruptions have also been associated with some large SEP events, although of softer spectrum (Kahler et al. 1986). Finally, we emphasize that flares are not exclusively an active region phenomenon. Two-ribbon flares can occur from quiescent filament regions (see, e.g., Fig. 1).

The flares considered by Hundhausen (1993) are generally confined to the sunspot latitudes because he compiled them from the Solar Geophysical Data that lists flare locations from H-alpha observations when available. However, if we define flares by soft X-ray enhancements, every filament eruption has such an enhancement, observed as PEAs (see McAllister et al. 1996 for a good example). In fact, Gopalswamy et al. (2010a) plotted the locations of all flares from GOES Soft X-ray Imager and found flare locations extending to latitudes above 60° for the period 2004–2007 (see their Fig. 10). On the other hand, when the locations of GOES soft X-ray flares with size >C3.0 (i.e., 3.0 x $10^{-6}$ Wm$^{-2}$) alone were plotted, the locations were confined to sunspot latitudes, clearly following the sunspot butterfly diagram.

Thus, there is really no clear separation between flare and filament eruption events. The average speeds of CMEs associated with the so-called filament eruption and flare events do differ (Gosling et al. 1976; Sheeley et al. 1999; Moon et al. 2002). Based on CME height-time profiles, MacQueen and Fisher (1983) had suggested that CMEs associated with prominence eruptions and flares may have different acceleration mechanisms. Another way to look at this is that the amount of free energy available in the source regions may be different, but the eruption mechanism may be the same. Fast CMEs (>1000 km/s) associated with quiescent filament eruptions are not uncommon: Song et al. (2013) reported on 13 eruptions from cycle 23, which they referred to as "flareless CMEs". As we noted above, such eruptions do have PEAs in soft X-ray and EUV, which become particularly clear when the intensity in a small area around the filament is monitored: the intensity gradually increases similar to other gradual flares, although the intensity is low.

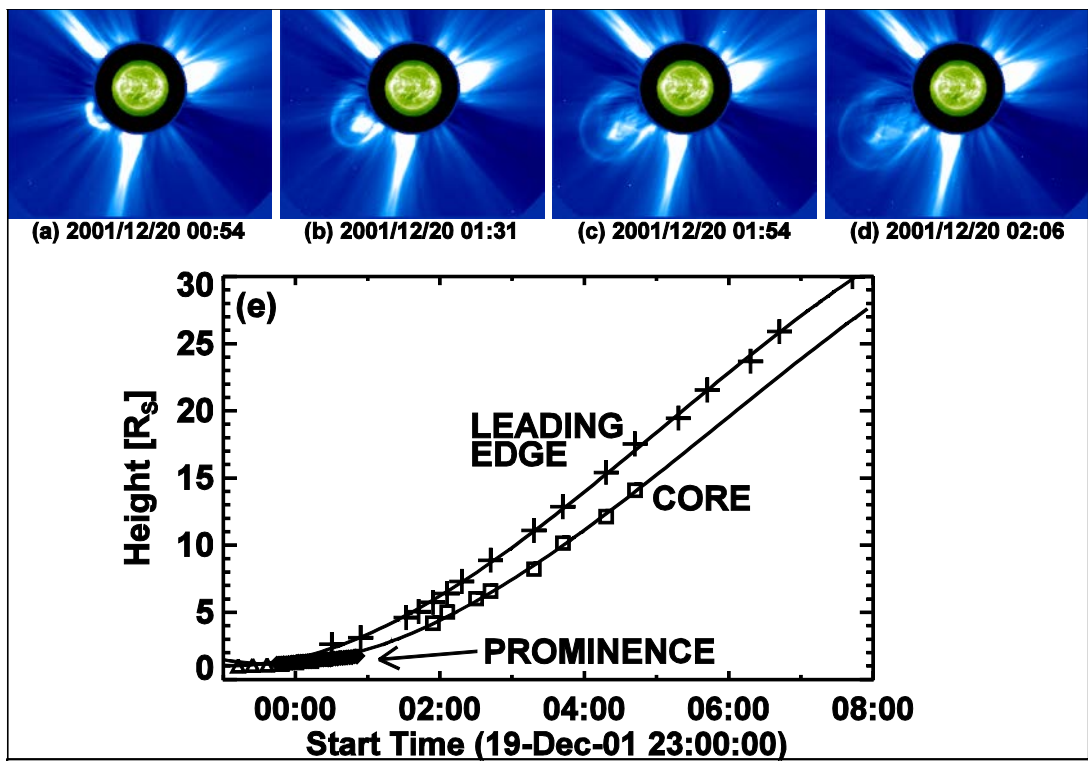

Figure 3. (top) A CME with three-part structure observed on 2001 December 20 by SOHO/LASCO. (bottom) Height-time measurements of the CME, prominence in the NoRH FOV and the prominence core in the LASCO FOV.

## 2.3 CME and PE kinematics

Eruptive prominences are observed as the brightest section of CMEs located in the interior of the CME structure. Soon after the discovery of white-light coronal mass ejections, it was realized that "analysis of eruptive prominence only or coronal mass ejection only would be incomplete without the other" (Schmahl and Hildner 1977). Comparing the kinematics of a CME and its prominence core, Webb and Jackson (1981) concluded that they moved out in a self-similar way. Figure 3 shows a recent example illustrating how the prominence core and the CME move together. The height-time plot shows the measurements close to the Sun made from NoRH images and then in the LASCO FOV. Within the NoRH FOV the prominence was still accelerating (~2.8 m/s$^2$) when it left the FOV and appeared as the CME core in the LASCO FOV. The CME was accelerating in the LASCO FOV (average acceleration ~14 m/s$^2$). The CME had an average speed of ~770 km/s in the LASCO FOV. The combined height-time plot shows that the prominence core closely followed the CME with a speed of ~660 km/s. Statistical analyses comparing the PE properties from below 2 Rs and CME properties in the coronagraphic FOV (above 2 Rs) for ~100 events yielded the following results (Gopalswamy et al. 2003a): 1. The CME core speed (average ~348 km/s) is always greater than the PE speed (average ~81 km/s) because of the continued acceleration. 2. Faster the PEs, the faster are the white-light cores. 3. The CME LE speeds are larger than the core speeds by ~43% (Maričić et al. 2009 found ~30% higher LE speed, but only for 18 events).

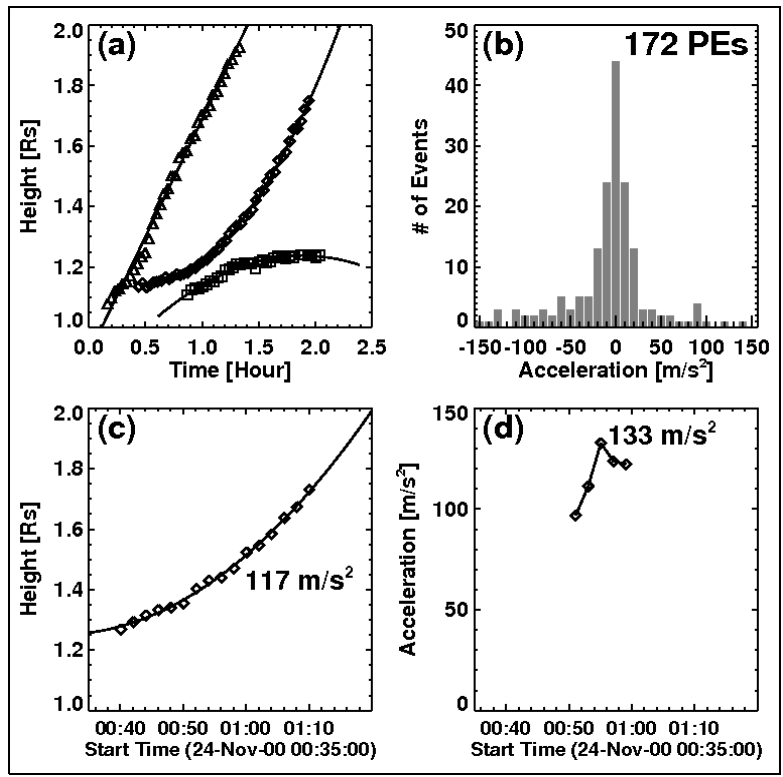

Figure 4. (a) Typical height-time profiles of eruptive prominences observed by NoRH. (b) Average accelerations derived from the height-time plots using quadratic fit). (c) The height-time profile of the 2000 November 24 prominence, which had an average acceleration of 117 m/s$^2$. (d) Time variation of the acceleration of the event in (c) when acceleration was computed taking 3–4 consecutive data points at a time.

The acceleration of PEs shows a lot of variations, as can be seen from the typical height-time plots shown in Fig. 4. The first two profiles in Fig. 4(a) in which the height continues to increase correspond to PEs that generally leave the Sun and become part of CME cores. The profile with decreasing height at later times corresponds to transverse PEs that do not get very far from the Sun. Tandberg-Hanssen et al. (1980) made a similar comparison between the height-time history of flare sprays and eruptive prominences. They pointed out the height-time plots are similar except that the initial acceleration phase is too quick to be observed. Engvold (1980) showed a number of height-time profiles representing the full range of accelerations and decelerations. Within ~2 Rs, the prominences have accelerations and decelerations as shown by the histogram in Fig. 4(b). CMEs with acceleration close to zero attain constant speed quickly, while those with positive acceleration continue to accelerate. PEs with a transverse trajectory and failed eruptions typically show deceleration. Deceleration is also observed when the end part of an eruption is captured; the material falling back shows a decrease in height with time. Fig. 4(c) shows the height-time profile of one of the PEs (2000 November 24) with a high acceleration in Fig. 4(b). This event also illustrates the quadratic fitting used in order to get the average acceleration values plotted in Fig. 4(b). The real acceleration is of course time dependent, as shown in Fig. 4(d) for the 2000 November 24 event. The maximum acceleration was ~133 m/s$^2$, only slightly higher than the average acceleration (~117 m/s$^2$). These values fall in the range of CME leading edge (LE) accelerations (Wood et al. 1999; Gopalswamy and Thompson 2000; Zhang et al. 2001; Zhang and Dere 2006; Vršnak et al. 2007; Maričić et al. 2009; Bein et al. 2011; Gopalswamy et al. 2012a). Maričić et al. (2009) showed that the CME LE acceleration was higher than that of the prominence core by a factor of ~2. The duration of the acceleration phase was about the same for the cores and CME LEs. The peak acceleration had an anti-correlation with the duration of acceleration for both components. The acceleration maximum was also attained around the same time for cores and LEs. The kinematic comparison between the CME core and the LE suggests that they evolve as a single structure moving away from the Sun. These observations are thus consistent with a flux rope with entrained cool material as a model for CMEs.

Nonthermal radio bursts that indicate plasma motion in the corona are closely associated with heated prominence material (Robinson 1978; Stewart et al. 1982; Gopalswamy and Kundu 1989). Imaging observations find that the moving type IV sources are located at the leading edge of eruptive prominences (see e.g., Gopalswamy and Kundu, 1989). The range of speeds derived from moving type IV bursts is roughly the same as that of prominence cores noted above (see also Robinson 1978). Since nonthermal electrons with energies of ~50 keV are needed to produce the moving type IV bursts, it is clear that particles accelerated during flare reconnection have access to the prominence structure and the surrounding flux ropes. In another case, nonthermal microwave emission was observed from the core and CME in a very fast event on 2001 April 18 (Gopalswamy 2006a). Hard X-ray emission from the prominence core was also observed in this event (Hudson et al. 2001). The heated plasma from flare reconnection also enters flux ropes, observed in situ as the high-charge state plasma.

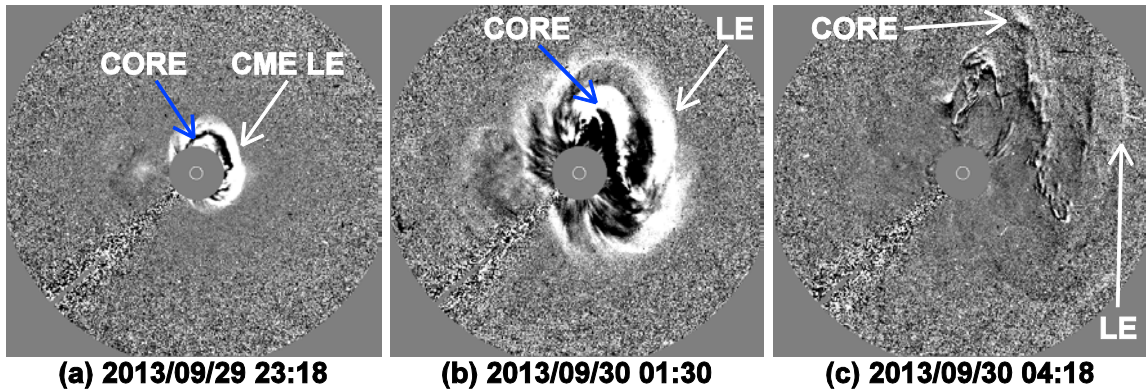

Figure 5. Three snapshots of the CME-prominence core system for the 2013 September 29 CME. In the image (c), part of the CME leading edge (LE) has left the LASCO FOV.

## 2.4 Prominences in the Heliosphere and Their Earth Impact

Even though significant amount material drains along the legs of eruptive prominences, one can track prominence cores readily to the edge of the LASCO FOV. In many cases, the cores retain their initial shape throughout the LASCO FOV. Figure 5 shows an eruptive prominence that became the core of the 2013 September 29 CME. The core maintained its shape all the way to the edge of the LASCO FOV (~32 Rs) and probably beyond. The north end of the core left the LASCO FOV at ~5 UT, while the south end moved out of the FOV ~ 7 h later.

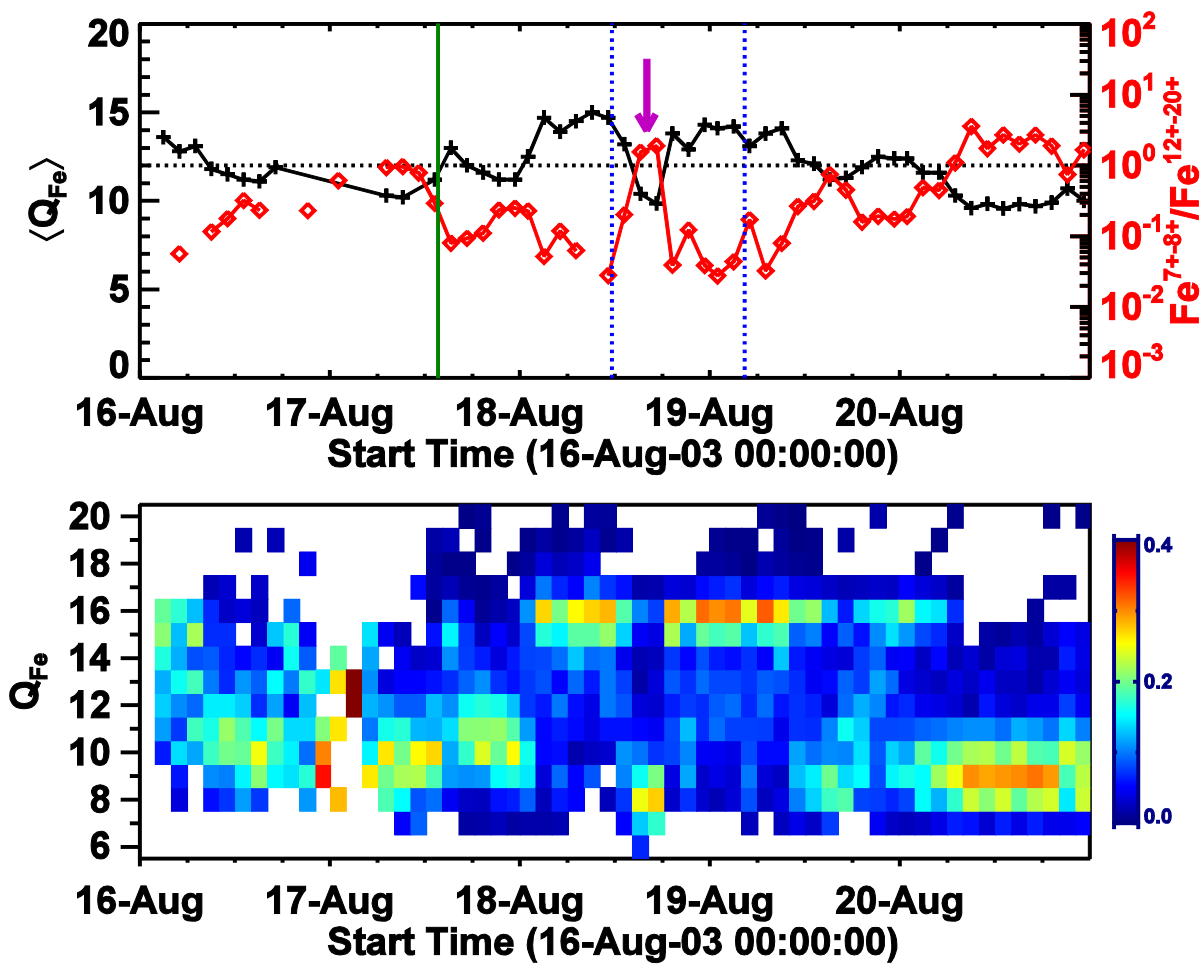

Figure 6. (top) Average Fe charge state (<$Q_{Fe}$>) and the ratio of low-to-high Fe charge states ($QFe^{7+-8+}/QFe^{12+-20+}$ ratio - red) during 17–19 August 2003 interplanetary CME (ICME) as observed by ACE/SWICS. The shock (green) and ICME (blue) times are marked. The narrow structure within the ICME (arrow) with low charge states is likely to be prominence material. (bottom) Individual ion charge state abundance (relative to the total abundance of Fe), from which the top curves were derived. Heavy ion charge states connect solar and in-situ observations.

Early observations indicated that prominence material remained at low temperatures to large distances. From H-alpha observations of a prominence core, Schmahl and Hildner (1977) reported that the core was at a temperature of only $\sim 2 \times 10^4$ K at a distance of $\sim 3$ Rs. In some cases, the filament gets heated to coronal temperatures much sooner (Webb and Jackson, 1981). When filaments erupt, the microwave brightness temperature typically increases to $\sim 10^4$ K from $\sim 8000$ K, and remains roughly the same near the Sun. This should make the filament disappear because the quiet Sun at 17 GHz has a brightness temperature of $\sim 10^4$ K (decreased contrast). However, when the heated "invisible" filament moved over a nearby plage, it obscured the plage for the duration of the transit of the filament over the plage (Hanaoka and Shinkawa, 1999). Similarly, Gopalswamy and Yashiro (2013) reported that a heated eruptive filament obscured the PEA of a nearby flare. These observations suggest that the core of the filament remains at $\sim 8000$ K but the outer sheath is heated to transition region temperatures ($\sim 10^5$ K). Since the sheath plasma is optically thin, it contributes only a few times 1000 K to the microwave brightness temperature, which explains the observed $10^4$ K. These observations suggest that a slow evaporation of the prominence occurs at least during the early phase of the eruption.

CMEs are observed throughout the heliosphere as flux ropes (see e.g. Richardson et al. 2006). One would certainly expect prominence material to be found inside the interplanetary CMEs (ICMEs). Prominence material is often observed at 1 AU inside interplanetary CMEs along with flare material (Burlaga et al. 1998; Gopalswamy et al. 1998; Lepri and Zurbuchen 2010; Gilbert et al. 2012; Gruesbeck et al. 2012). Figure 6 shows the high and low charge states within a magnetic cloud (MC) observed by Wind on 2003 August 19. The interval of elevated Fe charge states corresponds to the flare plasma. In the middle of the enhanced charge state region, there is a small interval (12–18 UT on August 18) where the average Fe charge state drops to +10. In order to further explore this interval, we examined the lower Fe charge states. To make it definitive, we compared the high ($\geq$+12) and low (+7 and +8) Fe charge states during the interval around the MC. The low-to-high Fe charge state ratio ($QFe^{7+-8+}/QFe^{12+-20+}$) exceeds 1 in the narrow interval where the Fe charge state dropped to +10 (see the bottom panel of Fig. 6). The charge state observations confirm the basic CME morphology: frontal structure, coronal void (flux rope), and prominence core; a shock in addition if the CME is fast (as is the case in Fig. 6).

Sharma and Srivastava (2012) reported a similar depression in ion charge states and elevated $He^+/He^{2+}$ ratio in intervals identified as filament material at the rear of two MCs. One of the MCs was from the rise phase of solar cycle 24 and the other from the declining phase of cycle 23, but they showed similar filament signatures. Identifying the filament material using elevated $He^+/He^{2+}$ ratio, Kozyra et al. (2013) reported that the filament material in the 2005 January 21 CME reached the magnetosphere, allowing the formation of a cold dense plasma sheet from within the magnetosphere from that material (see also Sharma et al. 2013; Dmitriev et al. 2014).

### 2.5 High Latitude Prominences and Prominence Eruptions

The latitude distribution of filaments is intimately connected to solar activity and hence important in understanding the long-term behavior of solar magnetism. In particular the high-latitude filaments that form the polar crown are of interest because they occur only during the maximum phase of solar cycles (see e.g., Ananthakrishnan 1952). Disappearance of the bipolar regions of the polar crown is essential for the sign reversal at the poles. This can happen when the polar crown filaments (PCFs) erupt as part of polar CMEs (Gopalswamy et al. 2003a), providing a way to track the PCFs without observing all of them (see below for an update). The polar CMEs are also important in understanding the eruption mechanism for CMEs because their source regions are purely bipolar regions. The PCF situation is similar to that of quiescent filament regions at lower latitudes. Thus CME eruption mechanisms applicable to low-latitude quiescent filament regions (e.g. Moore and Sterling, 2006) are likely to be valid for the polar CMEs also. The same mechanism may be applicable for active region CMEs as well because these regions also contain filaments overlying polarity inversion lines (see e.g., Vemareddy et al. 2012). Mechanisms that require multipolar configuration (e.g. Antiochos et al. 1999) may not apply to eruptions from bipolar regions.

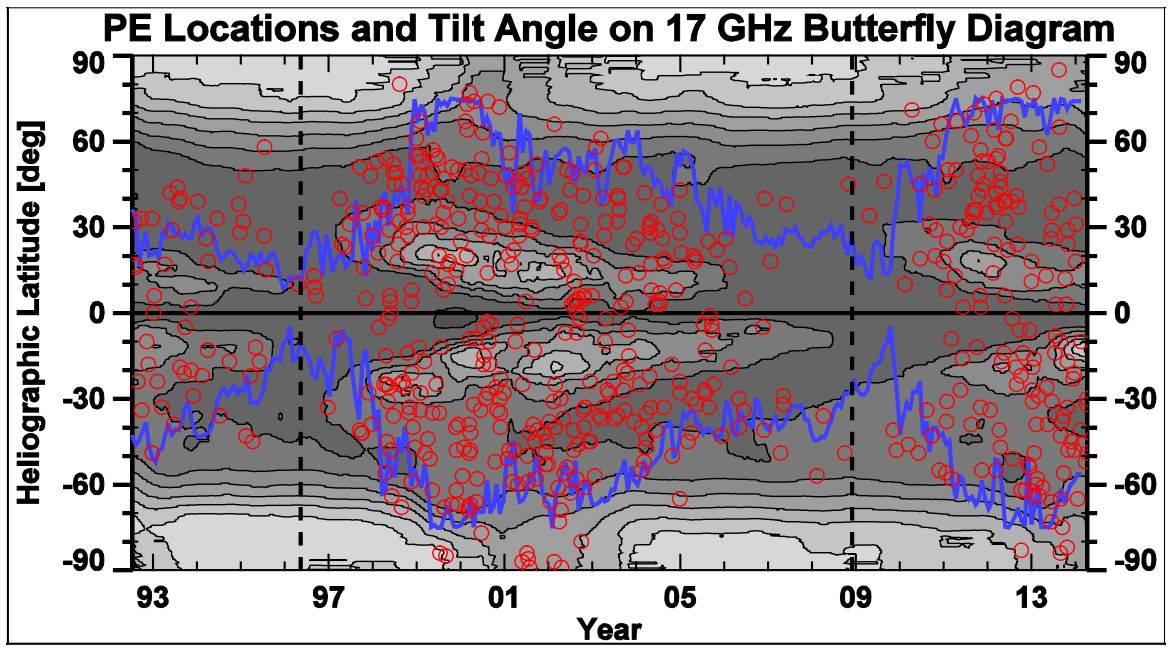

Figure 7. Several indicators of solar cycle phases. (1) The 17 GHz brightness temperature (contours) averaged longitudinally for each Carrington rotation and stacked together to see the time variation (the Microwave butterfly diagram). (2) The locations of PEs (red circles) detected automatically from NoRH images. These are limb events, so the projection effects are minimal and hence the eruption latitudes are known. (3) The tilt angle of the heliospheric current sheet (blue line) obtained from the Wilcox Solar Observatory. The vertical dashed lines denote the start times of cycles 23 (May 1996) and 24 (December 2008).

### 2.5.1. Signatures of the Rush to the Pole Phenomenon

The rush to the poles (RTTP) phenomenon refers to filaments that appear in the 40-50º latitude just before sunspot minimum and then systematically move toward the poles in both hemispheres (Lockyer 1931). RTTP was graphically demonstrated by Ananthakrishnan (1952) for the period from 1905 to 1950 (for cycles 14–18). Waldmeier (1960) and Hyder (1965) demonstrated the synchronism between the high-latitude filaments and the sign reversal at solar poles (see also Howard and Labonte 1981; Fujimori 1984; Lorenc et al. 2003; McIntosh 2003). The PCF disappearance lagged the reversal by several months, while the redevelopment of polar coronal holes (PCH) lagged by a few additional months.

Figure 7 illustrates the relationship among polar eruptions (PEs and CMEs), PCH and the tilt angle of the heliospheric current sheet. The distinct bright patches in microwaves at the poles (contour and gray-scale) correspond to PCH (Kosugi et al. 1986; Gopalswamy et al. 1999; 2012b; Shibasaki 2013). The polar microwave brightness enhancement is proportional to the polar magnetic field strength and corresponds to the chromosphere inside PCHs (Gopalswamy et al. 2012b). The disappearance of polar microwave emission corresponds to solar maximum phases. The low-latitude emission patches correspond to the active region emission (the microwave butterfly diagram). High tilt angles (>60º) correspond to the solar maximum phases. Note that the cycle 23/24 minimum is much extended and the onset of cycle 24 is delayed with respect to the time of peak polar brightness. In the north, the maxima of cycle 23 and 24 can be readily discerned. In the south, the cycle 23 maximum is bracketed by the end of cycle-22 maximum and the beginning of cycle-24 maximum. The locations of PEs red circles plotted on the chart are locations of prominence eruptions detected automatically from microwave images. PEs at latitudes >60° occur mainly during the maximum phase (indicated by high tilt angles), which is a representation of the RTTP phenomenon, except that we are tracking PEs rather than filaments or prominences. The cessation of PE activity at high latitudes marks the polarity reversal and the end of the maximum phase. There is clear north-south asymmetry in RTTP and polar sign reversal (see also Altrock 2014; Wang et al. 2002). In cycle 21 the PCF disappearance occurred first in the north, a trend that continued in cycles 22-24. Svalgaard and Kamide (2013) examined the hemispheric sunspot numbers since 1945 and concluded that the asymmetric polar sign reversal is a consequence of the hemispheric asymmetry in the sunspot activity: the hemisphere with dominant activity before the SSN maximum reverses first. Note that the sunspot asymmetry switched in cycle 20 (Svalgaard and Kamide, 2013), while the reversal asymmetry occurred in cycle 21. It is not clear why the switch in the reversal asymmetry happens with a lag of one cycle and what implications it may have for dynamo models (Leighton 1969).

### 2.5.2 Polar CMEs vs. Regular CMEs

Helmet streamers, coronal cavities, and filament channels are all related entities that define the pre-eruption environment of a prominence (see Engvold, 1989). Helmet streamers overlie cavities containing filaments as seen in eclipse pictures (e.g., Saito and Tandberg-Hanssen 1973). PCFs are no exception. When polar prominences appear at high latitudes, streamers can be found overlying them. Hansen et al. (1969) showed that the white-light brightness peak of the corona migrated from 50° to 80° during 1964-1967, consistent with the RTTP phenomenon (see Zhukov et al. 2008 for a polar streamer observed by SOHO/LASCO). In Kleczek's (1964) catalog of PEs, there were eruptions from various latitudes, including one from the polar zone that reached the largest height (>2 Rs).

Sheeley et al. (1980) were the first to report on a high-latitude CME observed by the Solwind coronagraph on board the P78-1 satellite on 1979 September 27. Sheeley et al. also speculated that there should be more such high-latitude CMEs citing the RTTP phenomenon. However, they were not able to find a solar source —neither a flare nor a prominence eruption—so they suggested that a change in the magnetic field configuration might have caused this CME. Sterling and Moore (2003) reported on a soft X-ray arcade from Yohkoh during the 1999 February 2 PCF eruption, although they did not study the CME association. Our examination revealed a relatively fast CME (average speed ~853 km/s). The acceleration was high (~60 m/s$^{-2}$) in the LASCO FOV so the speed exceeded ~1000 km/s before the CE left the coronagraph FOV. Details of this polar CME can be found in:
http://cdaw.gsfc.nasa.gov/CME_list/UNIVERSAL/1999_02/htpng/19990209.013005.p049s.htp.html.

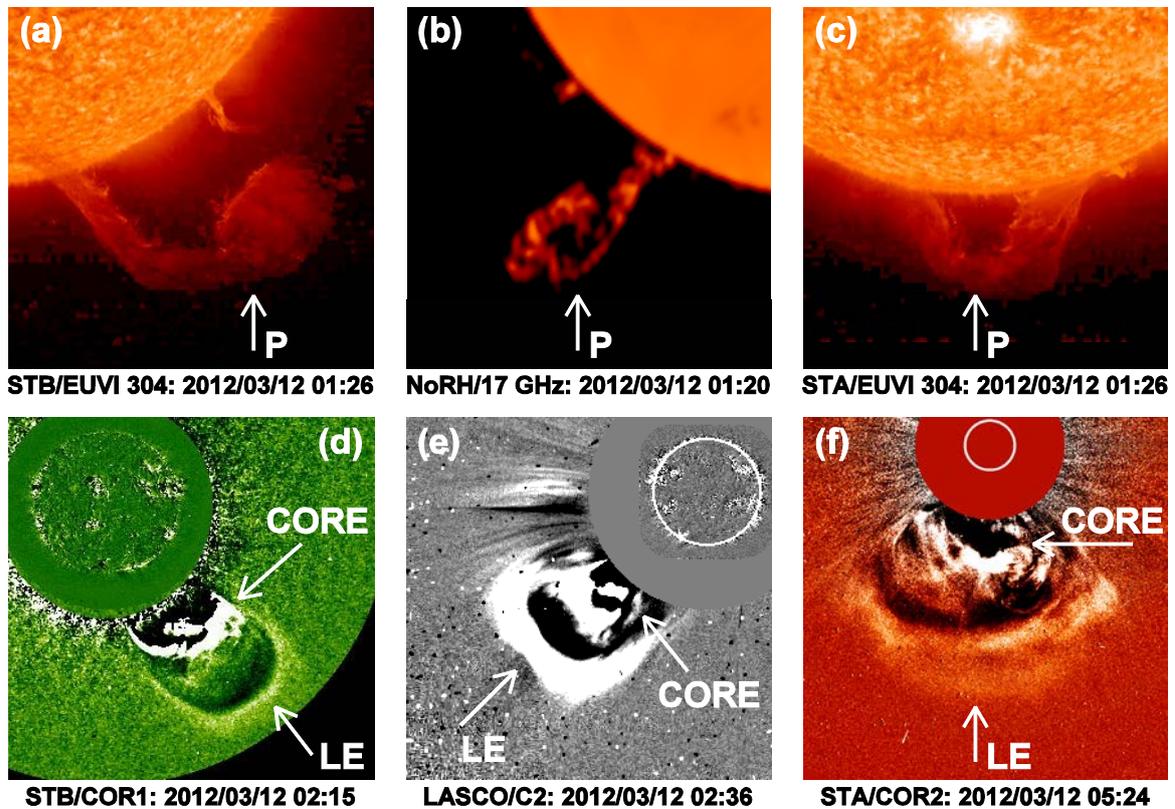

Figure 8. PCF eruption on 2012 March 12 (a-c) and the associated CME (d-f) from STEREO, NoRH and SOHO observations. The STEREO/EUVI 304 Å images show that it is truly from the polar crown. The prominence (P) becomes CME core in the outer corona as observed by STEREO/COR1 and LASCO/C2. Movies can be found in the CME catalog (Gopalswamy et al. 2009c) (http://cdaw.gsfc.nasa.gov/ CME_list/daily_movies/2012/03/12/).

The solar source of the 2012 March 12 CME is shown in Fig. 9. The PEA as observed by SDO/AIA (193 Å) is in the southeast quadrant because the filament extended beyond the east limb and appeared as a long east-west eruption in STB/EUVI (195 Å) FOV. The variations of the intensity (I) and its derivative dI/dt show the familiar pattern of gradual flares, except that the intensity is very low. The peak acceleration of the CME and core agree with the first dI/dt peak. The acceleration profile of the CME core is similar to that of the LE, but the magnitude is slightly smaller. The CME observation ended before the second peak. The I and dI/dt variations of the PEA are in good agreement between SDO and STB images. The acceleration of the CME LE peaked at ~200 m/s$^2$. The peak acceleration occurred when the CME LE was at a heliocentric distance of ~2.3 Rs, which is similar to the statistical value obtained by Bein et al. (2011) for a set of ~100 CMEs from low latitudes.

Gopalswamy (2013) showed that polar CMEs do have near-surface signatures such as two-ribbon flare structure and PEA. High-latitude eruptions started occurring in late 2010 for cycle 24, so we have scores of polar CMEs that can be compared with low-latitude CMEs (see Fig. 7). In addition to SOHO, we now have the Solar TErrestrial RElations Observatory (STEREO) and Solar Dynamic Observatory (SDO) observations to study polar CMEs and identify their solar sources unambiguously. One such PCF eruptions (2012 March 12) from Gopalswamy (2013) is shown in Fig. 8. The eruption occurred in the south polar zone near the east limb in Earth view as seen in the NoRH 17 GHz image. The prominence was also observed by SDO/AIA at 304 Å (not shown). At the time of the eruption, the STEREO-Behind (STB) spacecraft was located at E117. Therefore, the eruption was observed as a disk event close to the south pole in STB view. It was a backside event in STEREO-Ahead (STA) and the filament can be seen moving straight south in STA/EUVI 304 Å images. The CME was observed in the STB's inner coronagraph (COR1) FOV with a clear 3-part structure. The PEA formed at the initial location of the PCF as is evident from the STB/EUVI image superposed on the COR1 image. The CME appeared in the LASCO FOV at 01:26 UT and was observed until it crossed the FOV about half a day later. The CME was accelerating (~9 m/s$^2$) in the LASCO FOV and had an average speed of ~640 km/s. At the time of leaving the LASCO FOV, the CME had a speed of ~715 km/s. In the outer coronagraph (COR2) images, the CME is viewed broadsided and hence shows the full extent. The CME appearance is similar in STB/COR1 and COR2 images.

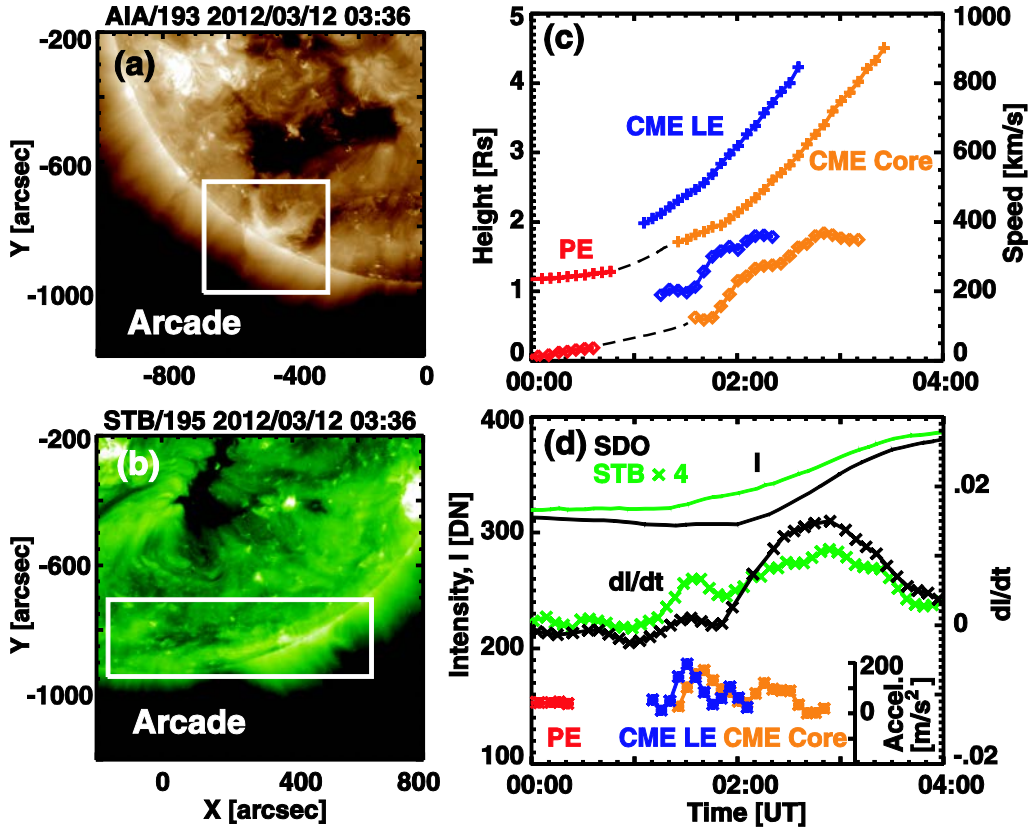

Figure 9. The post-eruption arcade (a,b) of the 2012 March 12 polar CME as imaged by SDO/AIA and STB/EUVI. The heights and speeds of the CMEs and the prominence core as measured from NoRH, SDO and STEREO images are in (c). The intensity I and its derivative dI/dt are compared with the acceleration profiles of the CME leading edge (LE) and the prominence core in (d). The intensity was computed as the average data number (DN) within the boxes drawn in (a) and (b).

There is another aspect of this eruption worth mentioning. The filament in the 2012 March 12 CME actually started rising towards the end of the previous day. SDO/AIA images taken before the filament rise show the lower part of the prominence cavity. Figure 10 shows the evolution of the cavity and prominence in three 171 Å SDO/AIA images. The LASCO/C2 images show the pre-eruption streamer overlying the cavity and prominence and two snapshots of the CME. The erupted cavity is seen in the LASCO images with the CME LE flattened. The fine thread-like feature near the outer edge of the cavity going over the prominence core of the CME is indicative of the flux rope structure. Such threads have been identified as field-line bundles that make up the flux rope (cavity) in white-light images (Chen et al. 1997). The flux rope was deformed somewhat between the 02:00 and 03:12 UT images. The cavity and prominence extended into the plane of the figure curving to the right as evidenced by the long east-west filament that erupted as observed by STB/EUVI images (see Fig. 10b). A similar prominence/cavity eruption was reported by Régnier et al. (2011), who concluded that the prominence is located at the bottom of the flux rope. It is now confirmed that cavities are ubiquitously found in quiescent filament regions and in the polar zone (Low and Hundhausen 1995; Gibson et al. 2010) and only occasionally in active regions. Detailed discussion on cavities and their relationship to prominences and CMEs can be found in Chapter 13.

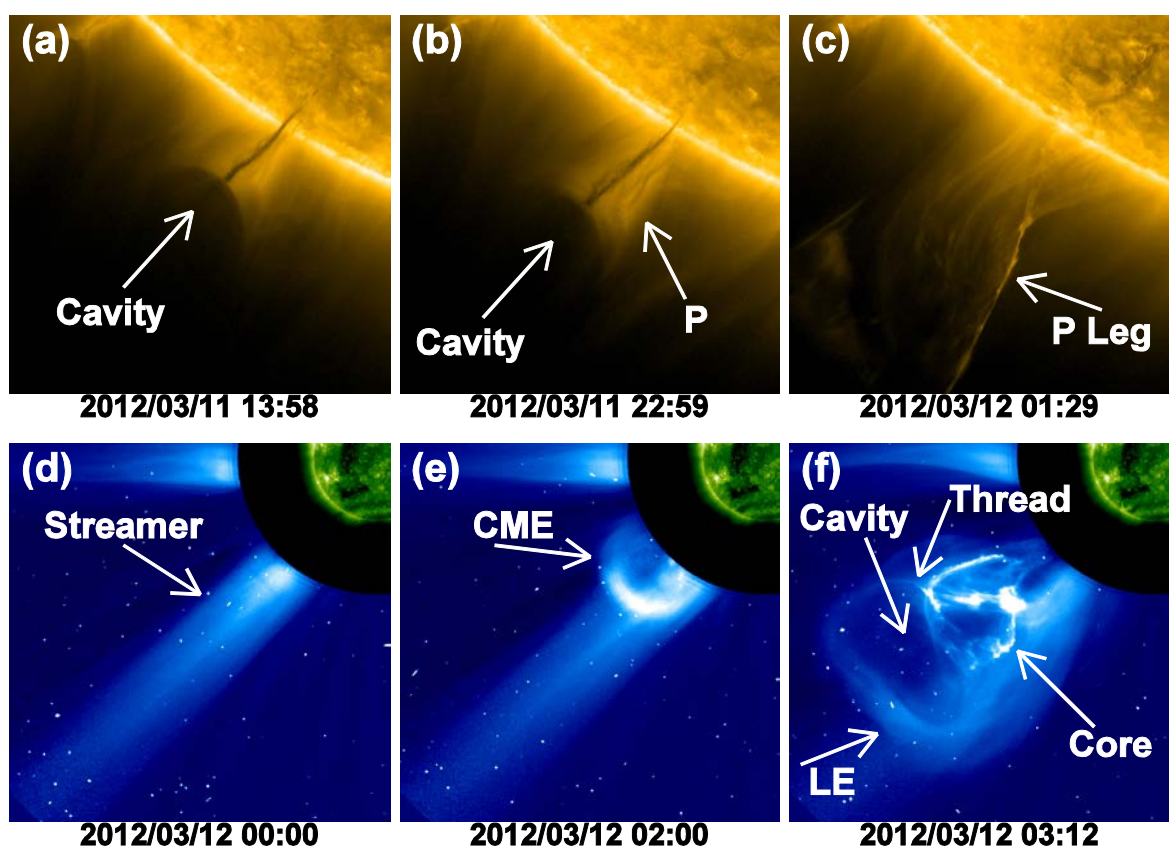

Figure 10. (top) SDO/AIA images at 171 Å showing the pre-eruption prominence and cavity (March 11, 13:58 UT), the slowly-rising cavity and prominence (March 11 22:59 UT) and the prominence leg after the cavity has left the FOV (March 12 01:29 UT). (bottom) Three LASCO/C2 images showing the polar streamer (March 12 00:00 UT), the early phase of the CME when the prominence core is still below the occulting disk (02:00 UT) and the whole CME with all the substructures: Leading edge (LE), cavity, and prominence core (03:12 UT). The fine thread that crosses the CME in the latitudinal direction is likely to be a bundle of field lines indicating the flux rope structure. The prominence core is the lateral section of the long filament that extends into the plane of the figure, curving to the right because it was observed so in STB view.

In summary, the polar CME of 2012 March 12 has all the classical features like any other CME associated with a prominence eruption demonstrating the following points. 1. The polar CMEs also have the three-part morphology. 2. The polar CME originates in a helmet streamer overlying the PCF. 3. A PEA is formed in each case, with its feet located on either side of the pre-eruption location of the filament (two-ribbon structure). 4. The CME speeds in the LASCO FOV are slightly above the average value (~475 km/s) of the general population of CMEs. 5. The peak value of the acceleration is typical of prominence associated CMEs (>100 m/s$^2$). 6. The CME attains the peak acceleration at a height of ~2.3 Rs, which is typical of most low-latitude CMEs.

Figure 11 further emphasizes the similarity between polar and low-latitude CMEs by showing the speed and width distributions. The CMEs were all associated with NoRH PEs listed in Gopalswamy et al. (2003b). The CMEs were divided into polar (latitude >60°) and low-latitude (≤40°) CMEs. CMEs in the latitude range 40-60° were omitted to make sure the events are truly from polar and low-latitude regions and not because of projection effects. The speeds and widths are quite similar suggesting that there is no significant difference between the two populations. The few higher-speed and wider CMEs are likely to be associated with active region filaments.

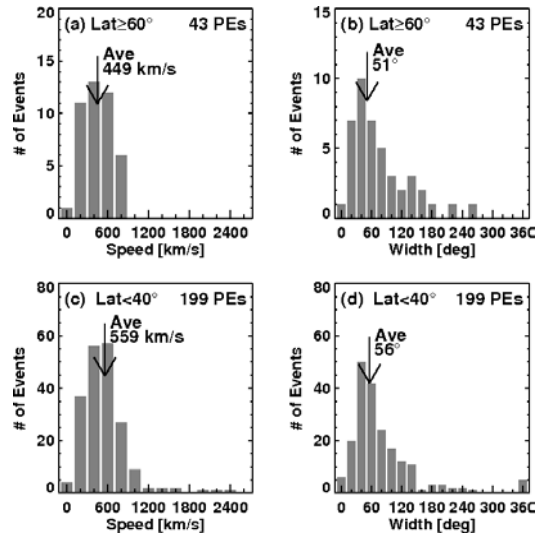

Figure 11. Speed and width distributions of cycle-23 polar CMEs (top) and low-latitude CMEs (bottom). Data from Gopalswamy et al. (2003b).

These observational facts confirm that the polar CMEs are similar to low-latitude CMEs, and contradict the suggestion that CMEs associated with polar crown filaments should not be considered as CMEs (Antiochos et al. 1999). These authors suggested that polar CMEs are similar to the blobs originating in helmet streamers at a heliocentric distance of about 3-4 Rs and accelerating slowly (~4 m/s$^2$) from ~150 km/s at 5 Rs to 300 km/s at ~25 Rs (Sheeley et al. 1997). Clearly, the peak acceleration of polar CMEs is larger by two orders of magnitude (see Fig. 9), unlike the Sheeley blobs, and similar to regular CMEs. Karpen et al. (2012) concluded that CMEs do occur even without flare reconnection, but the eruption will be slow, more like a streamer blowout (e.g., Sheeley et al. 1997) than a fast CME. They predicted a clear difference in the early acceleration profile between CMEs with and without impulsive flares. The observations presented here do indicate that the acceleration profile and magnitude are similar to that of regular CMEs (see also Joshi and Srivastava 2011). The polar CMEs are also associated with flare reconnection as evidenced by the PEAs. Thus polar CMEs carry mass, kinetic energy, and helicity from the source region into the interplanetary medium and remove the "PCF barrier" leading to the completion of the polarity reversal.

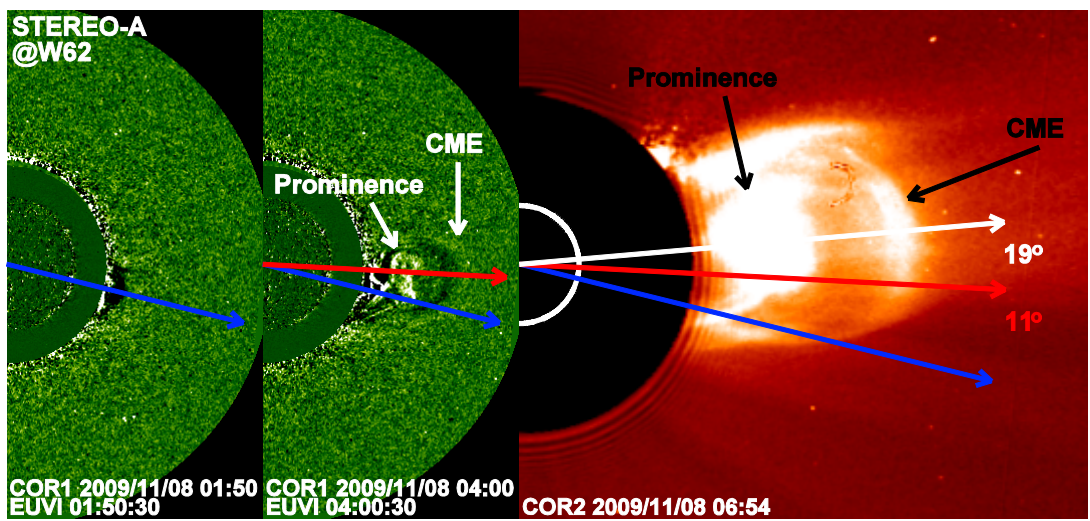

Figure 12. Deflection of the 2009 November 08 CME and prominence from the southwest initial direction (blue arrow) to the west (white arrow). The red arrow (11° away from the blue one) represents the CME position angle before it left the COR1 FOV. The images are from STA/EUVI, COR1, and COR2. Total deflection was by ~19° over a period of ~5 h.

### 3. Non-radial Motion of Eruptive Prominences and CMEs

A systematic equatorward deflection of CMEs observed during 1973–1974 by ~2° in the inner corona was reported by Hildner (1977), who concluded that (i) there must be a nonradial (equatorward) force acting on the CMEs and (ii) this must result in an enhanced effect of CMEs on the near-ecliptic IP medium. MacQueen et al. (1986) confirmed the deflection (average ~2°, maximum ~10°), but dismissed the possibility that it may have enhanced CME impact on the near-ecliptic IP medium. Decades of CME observations have confirmed the importance of CME deflection in understanding the propagation and geo-impact of CMEs.

Prominences were also found to have nonradial motion (Gopalswamy et al. 2000; Gopalswamy and Thompson, 2000; Simnett 2000). Gopalswamy and Thompson (2000) found that both the prominence and the CME showed the deflection, suggesting that the CME deflected as a whole. The initial position angle of the prominence eruption was 120° (or S30 in latitude). When the CME was observed in the LASCO/C3 FOV, both the CME and the prominence were at the position angle PA = 90° (at the equator). Thus the equatorward deflection was ~30°, which is the PA offset between the initial PE location and the CME nose.

Figure 12 illustrates a CME deflection event observed in full detail by the STEREO coronagraphs COR1 and COR2. The prominence erupted at a position angle of 256° at 01:50 UT. In the next two hours, the CME nose and the eruptive prominence had a significant movement toward the equator to PA~267°. Finally, the CME was near the equatorial plane (PA~275º) by 06:54 UT as observed by STEREO/COR2. The latitudinal movement of the CME can be quantified as 3°.75 per hour. This is typical of CMEs deflected toward equator in the rise phase of cycles 23 and 24. One of the consequences of this deflection is that relatively more magnetic clouds are observed near Earth during the rise phase of solar cycles resulting in intense geomagnetic storms (Gopalswamy et al. 2008).

Figure 13 shows the CME and prominence deflection in the 1998 January 25 event close to the Sun because SOHO's inner coronagraph LASCO/C1 was still operating. In the 14:54 UT frame, only the CME LE was observed. In the 15:14 UT frame, both the CME and the prominence core were visible. Between these two frames, the CME was deflected by ~12°. The CME was further tracked in the FOV of LASCO/C2 and C3, which indicated that the deflection ceased by ~16:00 UT, with a total deflection of ~ 17°. Figure 13 shows the PA of the CME nose as a function of time (t, measured in h from 13:30 UT). The e-folding time for trajectory change is ~1.2 h in the early phase.

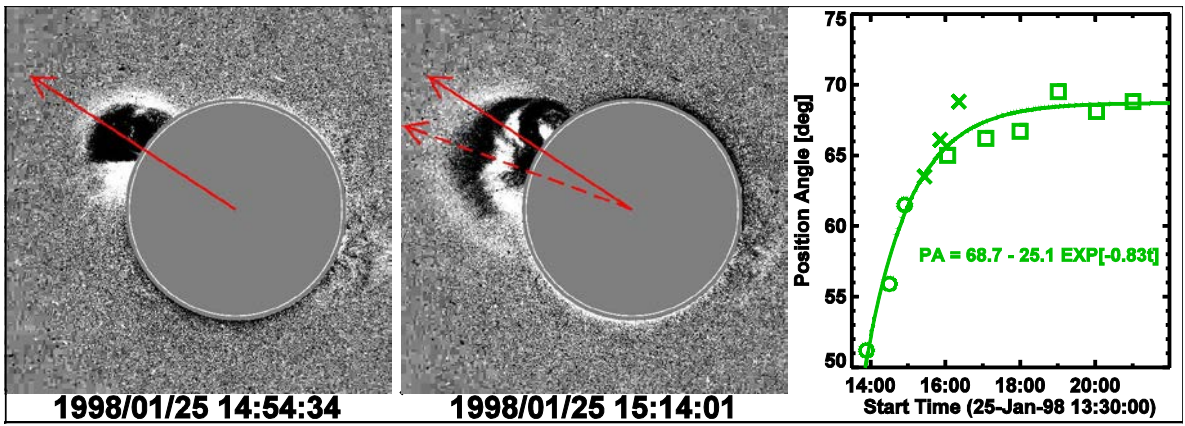

Figure 13. Two SOHO/LASCO/C1 images (left, middle) of the corona showing the 1998 January 25 CME deflecting toward the equator from a position angle of ~55° (solid arrow) to ~61° (dashed arrow). (right) Variation of the CME central position angle as a function of time. Circles, crosses, and squares represent measurements, respectively from C1, C2, and C3 coronagraphs of SOHO/LASCO. The solid line is the fit to the data points, showing that the nonradial motion stopped within a time T~2 h from the beginning of the eruption.

Plunkett et al. (2001) reported that the initial location of eruptions observed in EUV was also offset poleward of the associated CMEs. They considered 135 CMEs during April-December 1997 and found a bimodal distribution for eruption latitudes in EUV, while the corresponding CME latitudes were unimodal. These observations also indicated an average offset of ~30°. Gopalswamy et al. (2003b) investigated more than 200 PEs detected by NoRH, which revealed offsets as large as ~40° (see their Fig. 12). The offset was systematically poleward of the associated CMEs in the rise phase. The offset was not systematic during the maximum phase as was noted by MacQueen et al. (1986). The systematic offset found for the rise phase of cycle 23 is again confirmed in the rise phase of cycle 24 as shown in Fig. 14 (see also Gopalswamy et al. 2012b).

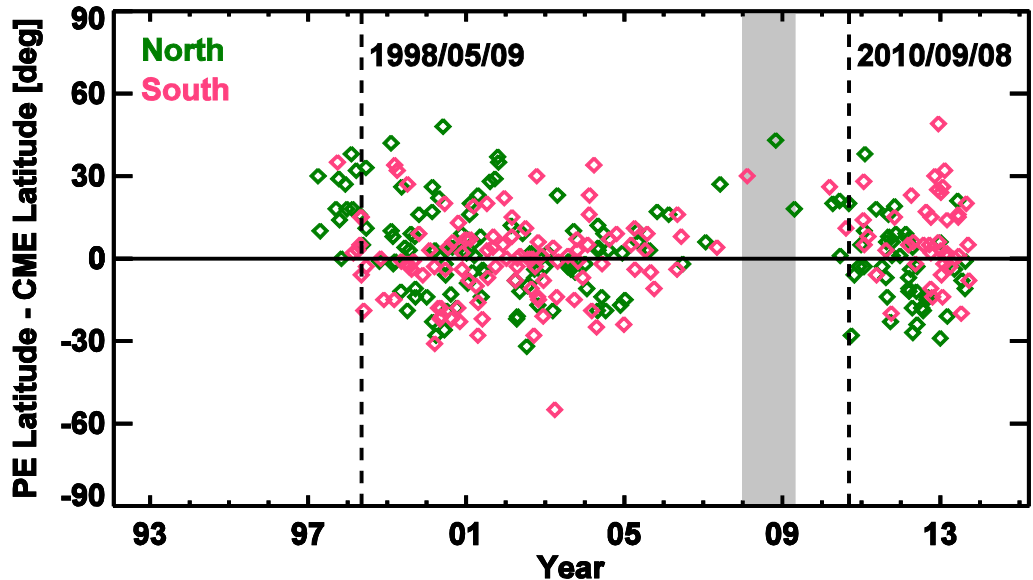

Figure 14. The latitude offset between PE (NoRH) and CME (LASCO). PEs originating in the northern and southern hemispheres are distinguished. The vertical dashed lines mark the times when the systematic poleward offset of CMEs with respect to the PEs ended. The shaded region corresponds to the time when cycles 23 and 24 overlapped (updated from Gopalswamy et al. 2012b).

Now we examine why Hildner (1977) and MacQueen et al. (1986) observed much smaller deflection. One possibility is that these authors measured the deflection in the narrow radial range of 2-3 Rs. In order to see if the deflection increases with radial distance, we have plotted the CME position angle as it moved out. Figure 15 shows the change in the PA of the CME nose as a function of the CME heliocentric distance (H) for two CMEs: the 1998 January 25 CME from cycle 23 (Fig. 14) and the 2009 May 5 CME from cycle 24. Clearly the PA changes until the CME reaches a certain height and then becomes stable. The PA vs. H curves have the form, $PA = A - B \exp(-H/H^*)$ where A and B are constants and $H^*$ is the e-folding distance. In the 1998 January 25 case, the maximum PA was attained when H was ~5 Rs ($H^*$ ~1.2 Rs). On the other hand, the deflection occurred over a much larger height and gradually ($H^*$ ~5 Rs) in the 2009 May 5 event.

The large deflections noted above support the original suggestion by Hildner (1977) that the deflection may have implications for the plasma in the equatorial region. The deflection towards the equator has been suggested as the reason for the relative higher rate of detection of magnetic clouds during the rise phase of solar cycle 23 (Gopalswamy, 2006b; Riley et al. 2006; Gopalswamy et al. 2008). All CMEs are likely to have a flux rope structure (i.e., magnetic cloud) and are observed so at 1 AU only when the observing spacecraft passes through the central part of the flux rope. The equatorward deflection during the rise phase thus allows CMEs to be detected as flux ropes at 1 AU.

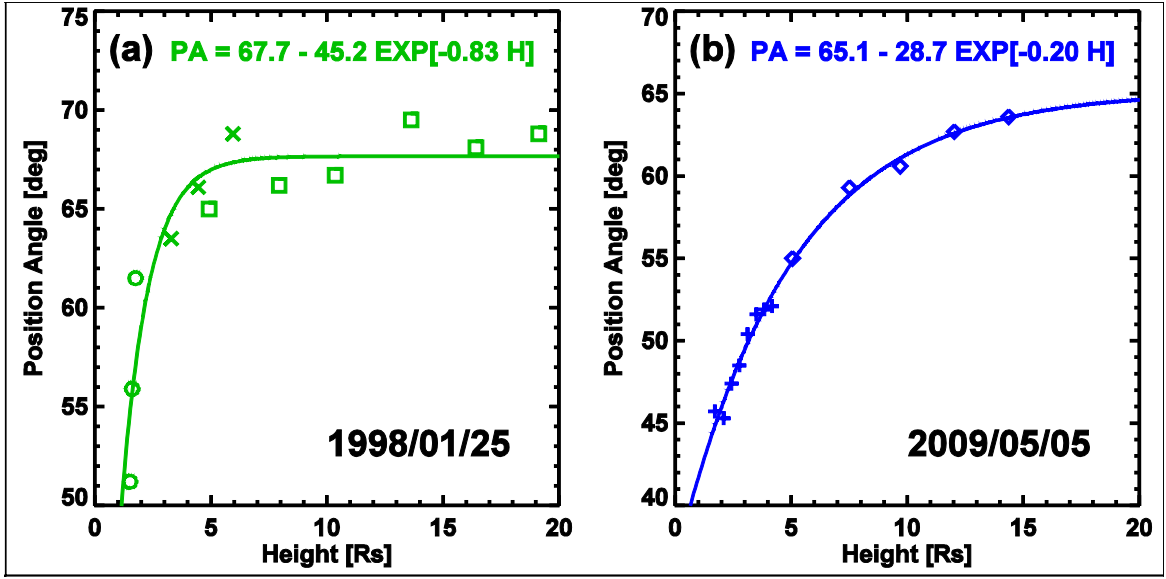

Figure 15. CME Position angle (PA) at various heliocentric distances (H). (a) The 1998 January 25 event with data points from SOHO/LASCO telescopes C1 (circles), C2 (crosses) and C3 (squares). (b) The 2009 May 5 event with data points from STEREO COR1 (crosses) and COR2 (diamonds). The solid curves are fits to the data points in the form PA = A - B exp (-H/H*), where A, B, and H* are coefficients of the fit given in the plots.

Both Hildner (1977) and MacQueen et al. (1986) attributed the nonradial motion to the global pattern of magnetic field and flow in the corona, which are distinct during solar minima and maxima. Filippov et al. (2001) proposed a simple axisymmetric model of the global magnetic field configuration with embedded flux rope to explain the nonradial motion of a prominence reported in Gopalswamy et al. (2000). During the minimum phase, PCHs are prominent and contain open magnetic fields of high strength. Active regions emerge at higher latitudes (~40°) during this phase, so CMEs erupt in the vicinity of the PCHs. The current thinking is that the magnetic field in coronal holes is responsible for CME deflection toward the equator. Coronal holes also occur at low latitudes, so similar deflection should happen when a CME erupts near low-latitude coronal holes. Gopalswamy et al. (2004b, 2005) reported coronal-hole influence on CMEs, with deflections toward and away from the Sun-Earth. Cremades et al. (2006) considered the area and distance to the eruption region of coronal holes (but not the magnetic field) to quantify the influence on CMEs. Gopalswamy et al. (2009a) introduced the photospheric magnetic field (B) inside coronal holes as another key parameter in determining the coronal hole influence on CMEs. They were able to show that many disk-center CMEs did not arrive at Earth because of deflection by large coronal holes located near the eruption region. However, the shocks associated with these CMEs did arrive at Earth and were called "driverless shocks" because the observing spacecraft did not intercept the CMEs. Gopalswamy et al. (2010b) found that $B^2$, rather than B, is a better representation of the force acting on the CMEs from the coronal holes. More quantitative investigations involve magnetic pressure gradient between the eruption region and the coronal hole (Gui et al. 2011; Panasenco et al. 2013; Kay et al. 2013), but the derived extent of deflection is similar to the previous work. Deflection from a high-field region can also occur in active regions, when the eruption occurs near large sunspots as described by Sterling et al. (2011b) for the 2006 December 13 CMEs. We also would like to point out that systematic nonradial motion can also be found in over-and-out CMEs that do not involve major PEAs (Moore and Sterling, 2007).

The deflections to the extent of 20–30° away from the Sun-Earth line are adequate to make the CMEs behave like limb CMEs and miss Earth (Gopalswamy et al. 2010b). Such driverless shocks were mostly observed during the declining phase of solar cycle 23, consistent with the abundance of low-latitude coronal holes in this phase. When the deflection is less severe, disk-center CMEs arrive at Earth as non-cloud ICMEs because the observing spacecraft pass through the edges of the flux ropes due to deflection by coronal holes (Gopalswamy et al. 2013; Xie et al. 2013; Mäkelä et al. 2013). So far we discussed the deflection of prominences and CMEs, which correspond to the main body of CMEs. Coronal holes can also deflect the shock surrounding the CMEs, which may be observed as reflected waves in EUV (Long et al. 2008; Gopalswamy et al. 2009b; Olmedo et al. 2012). Coronal hole deflection can also result in the lack of alignment between the ejecta and shock (Wood et al. 2012).

### 4. Implications for Models

The observations presented in this chapter indicate that eruptive prominences are an integral part of CMEs and typically occupy a small volume compared to the entire CME. Therefore, prominence eruptions need to be modeled along with the CME flux rope in which the prominence is embedded. There have been many observational signatures that confirm that almost all CMEs in the interplanetary medium have the flux rope structure. Therefore, the flux rope structure is fundamental to CMEs. When the flux rope is fast enough, it can drive a shock from certain distance in the corona. This has something to do with the interaction of the flux rope with the ambient medium. Observations of high and low charge states in flux ropes at 1 AU point to the intimate connection to the hot (flare) and cool (prominence) plasma originating near the Sun. The high charge state material is a characteristic of the IP flux rope irrespective of the primary near-surface feature (flare or prominence eruption). The approximate equality between flare reconnection flux and the azimuthal magnetic flux of IP flux ropes (Qiu et al. 2007) point to formation of flux ropes as part of the eruptive process. On the other hand, the presence of prominence cavity also indicates the presence of a flux rope, probably formed in a non-eruptive manner. The example shown in section 2.4 indicates that the cavity rises and leaves the Sun as a flux rope. This means an initial flux rope with some added flux seems to be the best possible scenario. Different proportions of existing and eruptive components of the flux are expected in different flux ropes in the IP medium.

The free energy that can be stored in closed magnetic regions on the Sun accounts for the observed range of CME speeds (<100 km/s to >3000 km/s) and accelerations (up to ~10 km/s$^2$). The acceleration can vary by two orders of magnitude from (100s of m/s$^2$ for eruptions from quiescent filament regions to >1 km/s$^2$ for active region eruptions). Thus the lowest CME acceleration is of the order of surface gravity of the Sun, which is negligible for the most energetic eruptions. Given the observed CME mass of up to $10^{16}$ g, these accelerations give an idea of the magnitude of the force involved in the eruptions.

The fact that polar CMEs behave like any other CMEs is an important factor that can help defining models. The polar CMEs originate from purely bipolar regions formed by the approaching of insurgent and incumbent fluxes in the polar zone. The bipolar nature of quiescent filament regions at low latitudes is no different. Many active regions also have bipolar structure and so produce CMEs. Models that require bipolar regions have the universal appeal, while those requiring multipolar configuration may work only in certain regions.

Finally propagation models need to account for deflections by other large-scale structures in the corona and IP medium. This is especially important for Earth-directed CMEs because the deflection can channel a CME toward or away from the Sun-Earth line. What is important is the deflection close to the Sun, which may not be properly taken into account by models such as ENLIL (Odstrcil and Pizzo 2009) whose inner boundary is at ~21 Rs. By the time CMEs reach this distance, the deflection effect might have disappeared.

**Acknowledgements**

I thank the editors, Oddbjørn Engvold and Jean-Claude Vial, for inviting this contribution. I thank David Webb, Pertti Mäkelä, Sarah Gibson, Ronald Moore, and Alphonse Sterling for helpful comments. I also thank S. Yashiro, S. Akiyama, and P. Mäkelä for their help with the figures. Work supported by NASA's LWS TR&T program.

**References**

Altrock RC (2014) Forecasting the Maxima of Solar Cycle 24 with Coronal Fe xiv Emission. Sol Phys 289:623-629

Ananthakrishnan R (1952) Prominence Activity and the Sunspot Cycle. Nature 170:156-158

Antiochos SK, Devore CR, Klimchuk JA (1999) A Model for Solar Coronal Mass Ejections. ApJ 510:485-493

Aulanier G (2014) The physical mechanisms that initiate and drive solar eruptions. In: Nature of Prominences and their role in Space Weather, B. Schmieder, J.-M Malherbe, S.T Wu (eds). Proceedings of the International Astronomical Union, IAU Symposium, vol 300, pp. 184-196

Bein BM, Berkebile-Stoiser S, Veronig AM et al. (2011) Impulsive Acceleration of Coronal Mass Ejections. I. Statistics and Coronal Mass Ejection Source Region Characteristics. ApJ 738:191-204

Burlaga L, Fitzenreiter R, Lepping R, Ogilvie K, Szabo A, Lazarus A et al. (1998) A magnetic cloud containing prominence material - January 1997. JGR 10:277

Chen J, Howard RA, Brueckner GE, Santoro R, Krall J, Paswaters SE, St. Cyr OC, Schwenn R, Lamy P, Simnett GM (1997) Evidence of an Erupting Magnetic Flux Rope: LASCO Coronal Mass Ejection of 1997 April 13. ApJ 490:L191-L194

Chen PF, Shibata K (2000) An Emerging Flux Trigger Mechanism for Coronal Mass Ejections. ApJ 545:524-531

Chifor C, Tripathi D, Mason HE, Dennis BR (2007) X-ray precursors to flares and filament eruptions. A&A 472:967-979

Cremades H, Bothmer V, Tripathi D (2006) Properties of structured coronal mass ejections in solar cycle 23. AdSR 38:461-465

Dmitriev AV, Suvorova AV, Chao J-K, Wang CB, Rastaetter L, Panasyuk MI, Lazutin LL, Kovtyukh AS, Veselovsky IS, Myagkova IN (2014) Anomalous dynamics of the extremely compressed magnetosphere during 21 January 2005 magnetic storm. JGR 119:877-896

Engvold O (1980) Energy and mass injected by flares and eruptive prominences. In: Solar and interplanetary dynamics. Dordrecht, D. Reidel Publishing Co. pp. 173-187

Engvold, O (1989) Prominence environment. In: Dynamics and structure of quiescent solar prominences. Dordrecht, Kluwer Academic Publishers, pp. 47-76.

Fan Y-H (2014) Magnetism and dynamics of prominences: MHD equilibria and triggers for eruption. Solar prominences, pp XX-YY, eds. J.-C. Vial & O. Engvold, Springer

Feynman J Martin SF (1995) The initiation of coronal mass ejections by newly emerging magnetic flux. JGR 100:3355-3367

Filippov BP, Gopalswamy N, Lozhechkin AV (2001) Non-radial motion of eruptive filaments. Sol Phys 203:119-130

Fujimori K (1984) Latitude distribution of solar prominences in the years 1975-1981. PASJ 36:189-190

Gibson SE (2014) Coronal cavities: observations and implications for the magnetic environment of prominences. Solar prominences, pp XX-YY, eds. J.-C. Vial & O. Engvold, Springer

Gibson SE, Kucera TA, Rastawicki D et al. (2010) Three-dimensional Morphology of a Coronal Prominence Cavity. ApJ 724:1133-1146

Gilbert HR, Holzer TE, Burkepile JT, Hundhausen AJ (2000) Active and Eruptive Prominences and Their Relationship to Coronal Mass Ejections. ApJ 537:503-515

Gilbert JA, Lepri ST, Landi E, Zurbuchen TH (2012) First Measurements of the Complete Heavy-ion Charge State Distributions of C, O, and Fe Associated with Interplanetary Coronal Mass Ejections. ApJ 751:20-27

Gopalswamy N, Kundu MR (1989) A slowly moving plasmoid associated with a filament eruption. Sol Phys 122:91-110

Gopalswamy N (2004) A Global Picture of CMEs in the Inner Heliosphere. In: The Sun and the Heliosphere as an Integrated System. G. Poletto, S. T. Suess (eds). Astrophysics and Space Science Library. Vol 317 Kluwer Academic Publishers, Dordrecht, The Netherlands, p.201

Gopalswamy N (2006a) Radio Observations of Solar Eruptions. Solar Physics with the Nobeyama Radioheliograph, Nobeyama Solar Radio Observatory, Nobeyama, pp.81-94

Gopalswamy N(2006b) Properties of Interplanetary Coronal Mass Ejections. SSRv 124:145-168

Gopalswamy N (1999) X-ray and Microwave Signatures of Coronal Mass Ejections. In: Solar Physics with Radio Observations, T. Bastian, N. Gopalswamy and K. Shibasaki (eds.), NRO Report No. 479, pp.141-152

Gopalswamy N (2013) Observations of CMEs and models of the eruptive corona. In: SOLAR WIND 13: Proceedings of the Thirteenth International Solar Wind Conference. AIP Conference Proceedings, Volume 1539, pp. 5-10

Gopalswamy N, Hanaoka Y, Kosugi T, Lepping RP, Steinberg JT, Plunkett S, Howard RA, Thompson BJ, Gurman J, Ho G, Nitta N, Hudson HS (1998) On the relationship between coronal mass ejections and magnetic clouds, GRL 25:2485

Gopalswamy N, Shibasaki K, Thompson BJ Gurman J, DeForest C (1999) Microwave enhancement and variability in the elephant's trunk coronal hole: Comparison with SOHO observations. JGR 104:9767-9780

Gopalswamy N, Hanaoka Y, Hudson HS (2000) Structure and Dynamics of the Corona Surrounding an Eruptive Prominence. AdSR 25:1851-1854

Gopalswamy N and Thompson BJ (2000) Early life of coronal mass ejections. JASTP 62:1457-1469

Gopalswamy N, Lara A, Yashiro S, Howard RA (2003a) Coronal Mass Ejections and Solar Polarity Reversal. ApJ 598:L63-L66

Gopalswamy N, Shimojo M, Lu W, Yashiro S, Shibasaki K, Howard RA (2003b) Prominence Eruptions and Coronal Mass Ejection: A Statistical Study Using Microwave Observations. ApJ 586:562-578

Gopalswamy N, Nunes S, Yashiro S, Howard RA (2004a) Variability of solar eruptions during cycle 23. AdSR 34:391-396

Gopalswamy N, Yashiro S, Krucker S, Stenborg G, Howard RA (2004b) Intensity variation of large solar energetic particle events associated with coronal mass ejections. JGR 109:A12105

Gopalswamy N, Yashiro S, Michalek G, Xie H, Lepping RP, Howard RA (2005) Solar source of the largest geomagnetic storm of cycle 23. GRL 32:L12S09

Gopalswamy N Mikić, Z Maia D, Alexander D, Cremades H, Kaufmann P, Tripathi D, Wang Y-M (2006) The Pre-CME Sun. SSRv 123:303-339

Gopalswamy N, Akiyama S, Yashiro S, Michalek G, Lepping RP (2008) Solar sources and geo-space consequences of interplanetary magnetic clouds observed during solar cycle 23. JASTP 70:245-253

Gopalswamy N, Mäkelä P, Xie H, Akiyama S, Yashiro S (2009a) CME interactions with coronal holes and their interplanetary consequences. JGR 114:A00A22

Gopalswamy N, Yashiro S, Temmer M, Davila J, Thompson WT, Jones S, McAteer RTJ, Wuelser J-P, Freeland S, Howard RA (2009b), EUV Wave Reflection from a Coronal Hole. ApJ 691:L123-L127

Gopalswamy N, Yashiro S, Michalek G, Stenborg G, Vourlidas A, Freeland S, Howard R (2009c) The SOHO/LASCO CME Catalog. Earth, Moon, and Planets 104 295-313

Gopalswamy N, Akiyama S, Yashiro S, Mäkelä P (2010a) Coronal Mass Ejections from Sunspot and Non-Sunspot Regions. In: Magnetic Coupling between the Interior and Atmosphere of the Sun, S.S. Hasan and R.J. Rutten (eds). Astrophysics and Space Science Proceedings. Springer Berlin Heidelberg. pp 289-307

Gopalswamy N, Mäkelä P, Xie H, Akiyama S, Yashiro S (2010b) Solar Sources of "Driverless" Interplanetary Shocks. In: TWELFTH INTERNATIONAL SOLAR WIND CONFERENCE. AIP Conference Proceedings 1216, pp. 452-458

Gopalswamy N, Xie H, Yashiro S, Akiyama S, Mäkelä P, Usoskin IG (2012a) Properties of Ground Level Enhancement Events and the Associated Solar Eruptions During Solar Cycle 23. SSRv 171:23-60

Gopalswamy N, Yashiro S, Mäkelä P, Michalek G, Shibasaki K, Hathaway DH, (2012b) Behavior of Solar Cycles 23 and 24 Revealed by Microwave Observations. ApJ 750:L42-L47

Gopalswamy N, Mäkelä P, Akiyama S, Xie H, Yashiro S, Reinard AA (2013) The Solar Connection of Enhanced Heavy Ion Charge States in the Interplanetary Medium: Implications for the Flux-Rope Structure of CMEs. Sol Phys 284:17-46

Gopalswamy N, Yashiro S (2013) Obscuration of Flare Emission by an Eruptive Prominence. PASJ 65:S11

Gopalswamy N, Akiyama S, Yashiro S, Xie H, Mäkelä P, Michalek G (2014) Anomalous expansion of coronal mass ejections during solar cycle 24 and its space weather implications. GRL 41:2673–2680

Gosling JT, Hildner E, MacQueen RM, Munro RH, Poland AI, Ross CL (1976) The speeds of coronal mass ejection events. Solar Phys 48:389-397

Gruesbeck JR, Lepri ST, Zurbuchen TH (2012) Two-plasma Model for Low Charge State Interplanetary Coronal Mass Ejection Observations. ApJ 760:141

Gui B, Shen C, Wang Y, Ye P, Liu J, Wang S, Zhao XP (2011) Quantitative Analysis of CME Deflections in the Corona. Sol Phys 271:111-139

Guo Y, Ding MD, Schmieder B, Li H, Török T, Wiegelmann T (2010) Driving Mechanism and Onset Condition of a Confined Eruption. ApJ 725:L38-L42

Hanaoka Y, Shinkawa T (1999) Heating of Erupting Prominences Observed at 17 GHz. ApJ 510:466-473

Hansen RT, Garcia CJ, Hansen SF, Loomis HG (1969) Brightness variations of the white light corona during the years 1964 67. Sol Phys 7:417-433

Hildner E, Gosling JT, Hansen RT, Bohlin JD (1975) The sources of material comprising a mass ejection coronal transient. Sol Phys 45:363-376

Hildner E (1977) Mass ejections from the corona into the interplanetary space. In: Study of Traveling Interplanetary Phenomena. M.A. Shea, D.F. Smart and S.T. Wu (eds), D. Reidel, Hingham, MA, pp. 3-21

Hori K, Culhane JL (2002) Trajectories of microwave prominence eruptions. A&A 382:666-677

Howard RA, Labonte, BA (1981) Surface magnetic fields during the solar activity cycle. Sol Phys 74:131-145

Hudson HS, Kosugi T, Nitta N, Shimojo M (2001) Hard X-Radiation from a Fast Coronal Ejection. ApJ 561:L211-L214

Hundhausen AJ (1993) Sizes and locations of coronal mass ejections - SMM observations from 1980 and 1984-1989. JGR 98:13177-13200

Hyder CL (1965) The "polar crown" of filaments and the Sun's polar magnetic fields. ApJ 141:271-273

Ji H, Wang H, Schmahl EJ, Moon Y-J, Jiang Y (2003) Observations of the Failed Eruption of a Filament. ApJ 595:L135-L138

Joshi AD, Srivastava N (2011) Kinematics of two eruptive prominences observed by EUVI/STEREO. ApJ 730:104-114

Kahler SW, Cliver EW, Cane HV, McGuire RE, Stone RG, Sheeley NR Jr (1986) Solar filament eruptions and energetic particle events. ApJ 302:504-510

Karpen JT, Antiochos SK, DeVore CR (2012) The Mechanisms for the Onset and Explosive Eruption of Coronal Mass Ejections and Eruptive Flares. ApJ 760:81-95

Kay C, Opher M, Evans RM (2013) Forecasting a Coronal Mass Ejection's Altered Trajectory: ForeCAT. ApJ 775:5-21

Kleczek J (1964) Occurrence of eruptive prominences. BAICz 15:41

Kosugi T, Ishiguro M, Shibasaki K (1986) Polar-cap and coronal-hole-associated brightenings of the sun at millimeter wavelengths. PASJ 38:1-11

Kozyra JU, Manchester WB, Escoubet CP, Lepri ST, Liemohn MW, Gonzalez WD, Thomsen MW, Tsurutani BT (2013) Earth's collision with a solar filament on 21 January 2005: Overview. JGR 118:5967-5978

Leighton RB (1969) A Magneto-Kinematic Model of the Solar Cycle. ApJ 156:1-26

Lepri ST, Zurbuchen TH (2010) Direct Observational Evidence of Filament Material Within Interplanetary Coronal Mass Ejections. ApJ 723:L22-L27

Liu K, Wang Y, Shen C, Wang S (2012) Critical Height for the Destabilization of Solar Prominences: Statistical Results from STEREO Observations. ApJ 744:168-177

Long DM, Gallagher PT, McAteer RTJ, Bloomfield DS (2008) The Kinematics of a Globally Propagating Disturbance in the Solar Corona. ApJ 680:L81-L84

Lorenc M, Pastorek L, Rybanský M (2003) Magnetic field reversals on the Sun and the N-S asymmetry. In: Solar variability as an input to the Earth's environment. Ed.: A. Wilson. ESA SP-535, Noordwijk: ESA Publications Division, pp. 129-132

Low BC, Hundhausen JR (1995) Magnetostatic structures of the solar corona. 2: The magnetic topology of quiescent prominences. ApJ 443:818-836

Lugaz N (2014) Eruptive prominences and their impact on the Earth: The impacts on our Earth and our life. Solar prominences, pp XX-YY, eds. J.-C. Vial & O. Engvold, Springer

Mackay DH, Gaizauskas V, Rickard GJ, Priest ER (1997) Force-free and Potential Models of a Filament Channel in which a Filament Forms. ApJ 486:534-549

MacQueen RM and Fisher RR (1983) The kinematics of solar inner coronal transients. Sol Phys 89:89-102

MacQueen RM, Hundhausen AJ, Conover CW (1986) The propagation of coronal mass ejection transients. JGR 91:31-38

Mäkelä P, Gopalswamy N, Xie H, Mohamed AA, Akiyama S, Yashiro S (2013) Coronal Hole Influence on the Observed Structure of Interplanetary CMEs. Sol Phys 284:59-75

Maričić D, Vršnak B, Roša D (2009) Relative Kinematics of the Leading Edge and the Prominence in Coronal Mass Ejections. Sol Phys 260:177–189

Marqué Ch, Lantos P, Klein KL, Delouis JM (2001) Coronal restructuring and electron acceleration following a filament eruption. A&A 374:316-325

Martin SF (1973) The Evolution of Prominences and Their Relationship to Active Centers (A Review). Sol Phys 31:3-21

McAllister AH, Dryer M, McIntosh P, Singer H, Weiss L (1996), A large polar crown coronal mass ejection and a ``problem'' geomagnetic storm: April 14-23. JGR 101:13497- 13516

McIntosh PS (2003) Patterns and dynamics of solar magnetic fields and HeI coronal holes in cycle 23. In: Solar variability as an input to the Earth's environment. A. Wilson (ed). ESA SP-535, Noordwijk: ESA Publications Division, pp. 807-818

Moon YJ, Choe GS, Wang H, Park YD, Gopalswamy N, Yang G, Yashiro S (2002) A Statistical Study of Two Classes of Coronal Mass Ejections. ApJ 581:694-702

Moore RL, Sterling AC, Hudson HS, Lemen JR (2001) Onset of the Magnetic Explosion in Solar Flares and Coronal Mass Ejections. ApJ 552:833-848

Moore RL, Sterling AC (2006) Initiation of Coronal Mass Ejections. In: Solar Eruptions and Energetic Particles, N. Gopalswamy, R. Mewaldt, J. Torsti (eds). Geophysical Monograph 165, AGU, Washington, DC, pp.43-57

Moore RL, Sterling AC (2007) The Coronal-dimming Footprint of a Streamer-Puff Coronal Mass Ejection: Confirmation of the Magnetic-Arch-Blowout Scenario. ApJ 661:543-550

Munro RH, Gosling JT, Hildner E, MacQueen RM, Poland AI, Ross CL (1979) The association of coronal mass ejection transients with other forms of solar activity. Sol Phys 61:201-215

Odstrcil D, Pizzo VJ (2009) Numerical Heliospheric Simulations as Assisting Tool for Interpretation of Observations by STEREO Heliospheric Imagers. Sol Phys 259:297-309

Olmedo O, Vourlidas A, Zhang, J, Cheng X (2012) Secondary Waves and/or the "Reflection" from and "Transmission" through a Coronal Hole of an Extreme Ultraviolet Wave Associated with the 2011 February 15 X2.2 Flare Observed with SDO/AIA and STEREO/EUVI, ApJ 756:143-155

Panasenco O, Martin SF, Velli M, Vourlidas A (2013) Origins of Rolling, Twisting, and Non-radial Propagation of Eruptive Solar Events. Solar Phys 287:391-413

Parenti S (2014) Solar Prominences: Observations. LRSP 11:1-88

Plunkett SP, Thompson BJ, St. Cyr OC, Howard RA (2001) Solar source regions of coronal mass ejections and their geomagnetic effects. JASTP 63:389-402

Qiu J, Hu Q, Howard TA, Yurchyshyn VB (2007) On the Magnetic Flux Budget in Low-Corona Magnetic Reconnection and Interplanetary Coronal Mass Ejections. ApJ 659: 758-772

Régnier S, Walsh RW, Alexander CE (2011) A new look at a polar crown cavity as observed by SDO/AIA Structure and dynamics. A&A 533:L1-L4

Reinard AA (2008) Analysis of interplanetary coronal mass ejection parameters as a function of energetics, source location, and magnetic structure. ApJ 682:1289

Richardson JD, Liu Y, Wang C, Burlaga LF (2006) ICMES at very large distances. AdSR 38:528-534

Riley P, Schatzman C, Cane HV, Richardson IG, Gopalswamy N (2006) On the Rates of Coronal Mass Ejections: Remote Solar and In Situ Observations. ApJ 647:648-653

Robinson RD (1978) Observations and Interpretation of Moving Type IV Solar Radio Bursts. Sol Phys 60:383-398

Saito K, Tandberg-Hanssen E, (1973) The Arch Systems, Cavities, and Prominences in the Helmet Streamer Observed at the Solar Eclipse, November 12, 1966. Sol Phys 31:105-121

Schmahl E, Hildner E (1977) Coronal mass-ejections-kinematics of the 19 December 1973 event. Solar Phys 55:473-490

Schmieder B, Démoulin P, Aulanier G (2013) Solar filament eruptions and their physical role in triggering coronal mass ejections. AdSR 51:1967-1980

Schrijver CJ, Elmore C, Kliem B, Török T, Title AM (2008) Observations and Modeling of the Early Acceleration Phase of Erupting Filaments Involved in Coronal Mass Ejections. ApJ 674:586-595

Sharma R, Srivastava N (2012) Presence of solar filament plasma detected in interplanetary coronal mass ejections by in situ spacecraft. JSWSC 2:A10

Sharma R, Srivastava N, Chakrabarty D, Möstl C, Hu Q (2013) Interplanetary and geomagnetic consequences of 5 January 2005 CMEs associated with eruptive filaments. JGR 118:3954-3967

Sheeley NR Jr, Howard RA, Koomen MJ, Michels DJ, Poland AI (1980) The observation of a high-latitude coronal transient. ApJ 238:L161-L164

Sheeley NR Jr, Wang Y-M, Hawley SH (1997) Measurements of Flow Speeds in the Corona between 2 and 30 Ro, ApJ, 484:472-478

Sheeley NR, Walters JH, Wang Y-M, Howard RA (1999) Continuous tracking of coronal outflows: two kinds of coronal mass ejections. JGR 104:24739–24768

Shibasaki K (2013) Long-Term Global Solar Activity Observed by the Nobeyama Radioheliograph. PASJ 65:S17-S22

Shimojo M (2013) Unusual Migration of Prominence Activities in the Southern Hemisphere during Cycles 23–24. PASJ 65:S16

Shimojo M, Yokoyama T, Asai A, Nakajima H, Shibasaki K (2006) One Solar-Cycle Observations of Prominence Activities Using the Nobeyama Radioheliograph 1992-2004. PASJ 58:85-92

Simnett GM (2000) The relationship between prominence eruptions and coronal mass ejections. JASTP 62:1479-1487

Song HQ, Chen Y, Ye DD, Han GQ, Du GH, Li G, Zhang J, Hu Q (2013) A Study of Fast Flareless Coronal Mass Ejections. ApJ 773:129-138

St. Cyr OC, Webb DF (1991) Activity associated with coronal mass ejections at solar minimum - SMM observations from 1984-1986. Sol Phys. 136:379-394

Sterling AC, Moore RL (2003) Tether-cutting Energetics of a Solar Quiet-Region Prominence Eruption. ApJ 599:1418-1425

Sterling AC, Moore RL, Freeland SL (2011a) Insights into Filament Eruption Onset from Solar Dynamics Observatory Observations. ApJ 731:L3

Sterling AC, Moore RL, Harra LK (2011b) Lateral Offset of the Coronal Mass Ejections from the X-flare of 2006 December 13 and Its Two Precursor Eruptions. ApJ 743:63-73

Stewart RT, Dulk GA, Sheridan KV, House LL, Wagner WJ, Illing R, Sawyer C (1982) Visible light observations of a dense plasmoid associated with a moving Type IV solar radio burst. A&A 116:217-223

Svalgaard L, Kamide Y (2013) Asymmetric Solar Polar Field Reversals. ApJ 763:23-28

Tandberg-Hanssen E, Martin SF, Hansen RT (1980) Dynamics of flare sprays. Sol Phys 65:357-368

Vemareddy P, Maurya RA, Ambastha A (2012) Filament Eruption in NOAA 11093 Leading to a Two-Ribbon M1.0 Class Flare and CME. Sol Phys 277:337-354

Waldmeier M (1960) Zirkulation und Magnetfeld der solaren Polarzone. Mit 7 Textabbildungen. Z Astrophys 49:176-185

Wang Y-M, Sheeley NR Jr (1999) Filament Eruptions near Emerging Bipoles. ApJ 510:L157-L160

Wang Y-M, Sheeley NR Jr, Andrews MD (2002) Polarity reversal of the solar magnetic field during cycle 23. JGR 107:SH10-1

Webb DF (2014) Eruptive prominences and their association with Coronal Mass Ejections. Solar prominences, pp XX-YY, eds. J.-C. Vial & O. Engvold, Springer

Webb DF, Jackson BV (1981) Kinematical analysis of flare spray ejecta observed in the corona. Sol Phys 73:341-361

Webb DF, Davis JM, McIntosh PS (1984) Observations of the reappearance of polar coronal holes and the reversal of the polar magnetic field. Sol Phys 92:109-132

Webb DF, Hundhausen AJ (1987) Activity associated with the solar origin of coronal mass ejections. Solar Phys 108:383-401

Webb DF, Howard RA (1994) The solar cycle variation of coronal mass ejections and the solar wind mass flux. JGR 99:4201-4220

Wood BE, Karovska M, Chen J, Brueckner GE, Cook JW, Howard RA (1999) Comparison of Two Coronal Mass Ejections Observed by EIT and LASCO with a Model of an Erupting Magnetic Flux Rope. ApJ 512:484-495

Wood BE, Wu CC, Rouillard AP, Howard RA, Socker DG (2012) A Coronal Hole's Effects on Coronal Mass Ejection Shock Morphology in the Inner Heliosphere. ApJ 755:43-52

Vršnak B, Maričić D, Stanger AL, Veronig AM, Temmer M, Roša D (2007) Sol Phys. 241, 85.

Xie H, Gopalswamy N, St. Cyr, O. C. (2013) Near-Sun Flux-Rope Structure of CMEs. Sol Phys 284:47-58

Zhang J, Dere KP, Howard RA, Kundu MR, White SM (2001) On the Temporal Relationship between Coronal Mass Ejections and Flares. ApJ 559:452-462

Zhang J, Dere KP (2006) A Statistical Study of Main and Residual Accelerations of Coronal Mass Ejections. ApJ 649:1100-1109

Zhukov AN, Saez F, Lamy P, Llebaria A, Stenborg G (2008) The Origin of Polar Streamers in the Solar Corona. ApJ 680:1532-1541